\documentclass[aps,showpacs, nofootinbib,floatfix]{revtex4}
\usepackage{graphicx, graphics, color}

\input amssym.tex
\input amssym.def

\newcommand{\be}{\begin{equation}}
\newcommand{\ee}{\end{equation}}
\newcommand{\ben}{\begin{eqnarray}}
\newcommand{\een}{\end{eqnarray}}

\newcommand{\la}{{\lambda}}

\newcommand{\na}{\nabla}

\newcommand{\Dsl}{{\slash \negthinspace \negthinspace \negthinspace \negthinspace  D}}

\newcommand{\talpha}{\tilde \alpha}

\newcommand{\ga}{\gamma}


\begin{document}

\title{Superconducting Hair on Charged Black String Background}

\author{{\L}ukasz Nakonieczny and Marek Rogatko}
\email{rogat@kft.umcs.lublin.pl, 
marek.rogatko@poczta.umcs.lublin.pl}
\affiliation{Institute of Physics \protect \\
Maria Curie-Sklodowska University \protect \\
20-031 Lublin, pl.~Marii Curie-Sklodowskiej 1, Poland }


\date{\today}

\begin{abstract}
Behaviour of Dirac fermions in the background of a charged black string penetrated by an Abelian Higgs
vortex is elaborated. One finds the evidence that the system under consideration can support fermion fields
acting like a superconducting cosmic string in the sense that a nontrivial Dirac fermion field can be carried by 
the system in question. The case of nonextremal and extremal black string vortex systems were considered. 
The influence of electric and Higgs charge, the winding number and the fermion mass on the fermion
localization near the black string event horizon was studied. It turned out that the extreme charged black string expelled
fermion fields more violently comparing to the nonextremal one.

\end{abstract}

\maketitle

\section{Introduction}
In the recent years there has been a great deal of attention paid to black object solutions in four and higher dimensions.
Among them black holes and black strings play the dominant role. 
For the first time uncharged and rotating black string solutions
were considered in \cite{lem95}. On the other hand, the generalization comprising the charged case was provided in 
Ref.\cite{lem96}.
Rotating charged black strings in dilaton gravity being the low-energy limit of the string theory 
with non-trivial potential were elaborated in \cite{deh05},
while the thermodynamics of the aforementioned objects was studied in \cite{deh02}.
\par
At the beginning of our Universe could undergo several phase transitions
which might produce stable topological defects like cosmic strings, monopoles and domain walls \cite{vil}.
Among them, cosmic strings and cosmic string black hole
systems acquire much interest. At the distributional mass source limit metric of this system was derived in \cite{ary86} (the so-called
{\it thin string limit}). In Refs.\cite{ach95,cha98} more realistic cases of an Abelian Higgs vortex 
on Schwarzschild and charged Reissner-Nordstr\"om (RN) black holes were elaborated.
It was revealed that an analog of {\it Meissner effect} (i.e., the expulsion of the vortex fields from the black hole)
could take place. It happened that this phenomenon occurred for some range of black hole parameters \cite{bon98}.
In the case of the other topological defects, like domain walls, the very similar phenomenon 
has been revealed \cite{domainwalls}.
\par
As the uniqueness theorem for black hole in the low-energy string theory was quite well established \cite{uniqth}
the next step in the aforementioned research was to consider Abelian Higgs vortex in the background of dilaton and
Euclidean dilaton black holes \cite{mod98a}-\cite{mod99}. On the contrary to the extremal black hole in Einstein-Maxwell
theory it happens that extremal dilaton black holes always expel vortex Higgs fields from their interior \cite{mod99}.
\par
On the other hand, studies of a much more realistic fields than scalar attracted more attention.
Solutions of field equations describing fermions in a curved
geometry is a challenge to the investigations of the underlying structure of the spacetime.
The better understanding of properties of black holes also acquires examination of the behaviour
of matter fields in the vicinity of them \cite{chandra}.  
Dirac fermions behaviour was studied in the context of Einstein-Yang-Mills background \cite{gib93}.
Fermion fields were analyzed in the near horizon limit of an extreme Kerr black hole \cite{sak04}
as well as in the extreme RN case  \cite{loh84}.
It was also revealed \cite{fin00}- \cite{findir}, that 
the only black hole solution of four-spinor Einstein-dilaton-Yang-Mills equations
were those for which the spinors vanished identically outside black hole.
Dirac fields were considered in Bertotti-Robinson spacetime \cite{br1,br2}
and in the context of a cosmological solution with
a homogeneous Yang-Mills fields acting as an energy source \cite{gib94}.
\par
A different issue, related to the problem of black hole uniqueness theorem, is the late-time decay of 
fermion fields in the background of various kinds of black holes.
The late-time behaviour of massless and massive Dirac fermion fields were widely studied
in spacetimes of static as well as stationary black holes \cite{jin04}-\cite{goz11}. 
\par
Fermion fields were also considered in the context of
brane models of our Universe, represented as $(3 + 1)$-dimensional submanifold living in higher dimensional
spacetime. Decay of massive Dirac 
hair on a brane black hole was considered in \cite{br08}.
\par
A direct consequence of the implementation of
fermions in the cosmic string theory caused that they become superconducting. 
On their own, superconducting cosmic strings might be responsible for various {\it exotic } astrophysical phenomena
such as the high-redshift gamma ray bursts \cite{che10}, the ultra-high energy cosmic rays (UHECRs)
\cite{nag00}.\\
In Ref.\cite{gre92} it was revealed that the Dirac operator in the 
spacetime of the system composed of
Euclidean magnetic RN black hole and a vortex in the theories containing superconducting 
cosmic strings \cite{wit85} possessed zero modes. In turn, the aforementioned zero modes caused the fermion
condensate around magnetic RN black hole. The generalization of the aforementioned researches to the case
of Euclidean dilaton black hole superconducting string system was provided in Ref.\cite{nak11}, where 
the non-zero Dirac fermion modes and an arbitrary coupling constant between
dilaton and Maxwell field were taken into account. It was found that Euclidean dilaton black hole spacetime could support
the superconducting cosmic string as hair.
\par
As far as the black string vortex system was concerned, 
the problem of an Abelian Higgs vortex solution in the background of charged black string was studied in Ref.\cite{deh02v}.
It was found that it could support the vortex as hair. The effect of the presence of the vortex 
was to induce a deficit angle in the spacetime of the static charged black string.
\par
Recently the gravity-gauge theory duality attracted a lot of attention. Due to the AdS/CFT correspondence gravity theory
in $d$-dimensional anti-de Sitter (AdS) spacetime is equivalent to a conformal field theory (CFT) on $(d-1)$-dimensional
spacetime which constitutes the boundary of the AdS manifold. In the context of the AdS/CFT correspondence the black 
strings played their important roles. Namely, in Ref.\cite{bri10} 
the uncharged rotating black strings were performed to describe a holographic fluid/superfluid phase transition.
On the other hand, the formation of scalar hair which corresponds to a holographic fluid/superfluid phase transition
and the formation of scalar hair on the tip of the solitonic, cigar-shaped solution describing a holographic insulator/
superfluid phase transition was considered in Ref.\cite{bri11}.
\par
Motivated by the above arguments, 
in our paper we shall investigate the problem of fermionic superconductivity in the spacetime of a charged black 
string pierced by an Abelian Higgs vortex. To our knowledge, the problem of Dirac superconductivity in the background of 
charged black string Abelian vortex system has not been elaborated before. In what follows we shall consider the
near-horizon behaviour of fermionic fields which are responsible for the superconductivity in the case of
nonextremal and extremal black string vortex system. The special attention we pay to the extremal black string case, due to 
the suspicions of the so-called {\it Meissner effect}, i.e., expulsion of the vortex fields from the black string interior.
Such a phenomenon took place in the extremal black hole vortex background and was studied previously.
\par
The layout of our paper will be as follows.
In Sec.II, for the readers convenience, we quote some basic facts concerning with the charged black string Abelian 
Higgs vortex configuration. Sec.III will be devoted to the fermionic superconductivity in the background of
charged black string vortex system. We derive Eqs. of motion for the fermionic fields and describe the spinors which 
are eigenstates of $\ga^0~\ga^3$ matrices. In Sec.IV we shall tackle the problem of the asymptotic behaviour of the 
fermion fields in question. In Sec.V we elaborate nonextremal charged black string and the behaviour of electrically charged 
and uncharged fermion fields and their influence on superconductivity. Sec.VI will be connected with the extremal charged black 
string and possibilities of  fermion condensation near its event horizon. In Sec.VII we conclude our researches.

\section{Charged Black String/Abelian Higgs Vortex Configuration }
In this section we shall discuss a charged black string/vortex configuration. In our studies we assume the 
complete separation of the degrees of freedom of the object under consideration. The charged black string line element will
be treated as the background solution on which one builds an Abelian Higgs vortex. The action governing the black 
string/ Abelian vortex is provided by the following expression:
\be
S = S_{1} + S_{bos},
\ee
where $S_{1}$ is 
the Einstein-Hilbert action for gravity with negative cosmological constant $\Lambda$ and $U(1)$-gauge field.
The corresponding gauge field can be thought as the everyday Maxwell field. $S_{1}$ action yields
\be
S_{1} = \int \sqrt{-g}~d^4 x
\left [ R - 2\Lambda  - F_{\mu \nu}F^{\mu \nu}
 \right], 
\label{acbs}
\ee
where $F_{\alpha \beta} = 2\na_{[ \alpha}A_{\beta ]}$.
The other gauge field is hidden in the action $S_{bos}$ and it is subject to the spontaneous symmetry
breaking. Its action implies
\be
S_{bos} = \int \sqrt{-g}~d^4 x
\left [
- (d_{\mu} \Phi)^{\dagger}d^{\mu}\Phi - \frac{1}{4} {B}_{\mu \nu}
 {B}^{\mu \nu}
- \frac{\lambda}{4} \bigg(\Phi^{\dagger}\Phi - \eta^2 \bigg)^2  \right],
\ee
where $B_{\mu \nu} = 2\na_{[\mu }B_{\nu ]}$ is the field strength associated with $B_{\mu}$-gauge field,
$\eta$ is the energy scale of symmetry breaking and $\la$ is the Higgs coupling.
The {\it covariant derivative} has the form
$d_{\mu} = \nabla_{\mu} + i~e_{R} B_{\mu}$, where $e_{R}$ is the gauge coupling constant.
\par
Consideration of a nonlinear system coupled to gravity constitutes a very difficult problem. 
However, it was shown that the self-gravitating Nielsen-Olesen vortex can act as a long hair.
The same situation takes place in the case of a charged black string \cite{deh02v}. \\
In what follows we choose the vortex fields $X$ and $P$ in the forms provided by the following expressions:
\be
\Phi(x^{i}) = \eta X(r)e^{iN \phi},
\label{phi}
\ee
and
\be
B_{\mu}(x^{i}) = \frac{1}{e_{R}} \left[P_{\mu}(r) - N \na_{\mu}\phi \right ] ,
\label{b}
\ee
where $\phi$ is the usual angular coordinate in the cylindrical spacetime.
Consequently, the bosonic action in terms of X and P yield
\be
S_{bos} = \int \sqrt{-g} d^4 x
\big [  - \eta^2 \nabla_{\mu}X \nabla^{\mu}X 
- \eta^2 P_{\mu}P^{\mu}X^2
- \frac{1}{4 e_{R}^2} \tilde{B}_{\mu \nu}\tilde{B}^{\mu \nu}  
- \frac{\lambda \eta^2}{2} (X^2 - 1)^2 \big ],
\ee
where $\tilde{B}_{\mu \nu} = \nabla_{\mu} P_{\nu} - \nabla_{\nu} P_{\mu}$.
Equations of motion for bosonic part of the action read
\ben
\nabla_{\mu} \nabla^{\nu}X &-& P_{\mu}P^{\nu}X - \frac{ \lambda \eta^2}{2} X(X^2 - 1) = 0, \\
\nabla_{\mu} \tilde{B}^{\mu \nu} &-& 2 \eta^2 e_{R}^2 P^{\nu} X^2 = 0. 
\een
A static cylindrically symmetric line element of a charged black string being subject to the action
(\ref{acbs}) takes the form \cite{lem96}
\be
ds^2 = - A^2 dt^2 + \frac{dr^2}{A^2} + r^2 d \phi^2 + \frac{r^2}{l^2} dz^2, 
\ee
where by $A^2$ we have denoted the following:
\be
A^2 = \frac{r^2}{l^2} - \frac{b l}{r} + 
\frac{\tilde{\lambda}^2 l^2}{r^2}.
\ee
The parameters $b,~\tilde{\lambda},~l$ are related to the black string
mass, charge per unit length and cosmological constant by the relations as follows:
\be
M = \frac{b}{4}, \qquad
Q_{BS} = \frac{\tilde{\lambda}}{2}, \qquad
\Lambda = - \frac{3}{l^2}.
\ee 
The gauge field component is chosen to be equal to $A_{\mu} = - \frac{l \tilde{\lambda}}{r} \delta^{t}_{\mu}$.
In the case when $- \infty < z < \infty$ the above line element describes a black string with cylindrical event horizon,
the inner $r_{-}$ and the outer $r_{+}$. For the specific value of $b$ parameter equals to
\be
b_{crit} = 4 \bigg( \frac{ \tilde{ \lambda }  }{ \sqrt{ 3 } } \bigg)^{ \frac{ 3 }{ 2 } },
\ee
one has the case when the outer and inner horizons of black string coincide, i.e., $r_{-} = r_{+}$.
Then, we obtain an extremal charged black string.
\par
As was mentioned, treating a nonlinear system coupled to gravity is very difficult and nontrivial problem. In order to 
circumvent the difficulties one can take a background solution and in this spacetime solve equations of motion for Higgs fields.
Due to the symmetry of the problem in question, let us consider quite general form of the cylindrically spacetime 
given by the line element
\be
ds^2 = - A(r)^2 dt^2 + B(r)^2 dr^2 + C(r)^2 d \phi^2 + D(r)^2 dz^2.
\label{lelem}
\ee
Taking into account the vortex field $\Phi$ (\ref{phi})
and the $\phi$-component of the gauge field $B_{\mu}$~(\ref{b}) , as well as 
and the line element
(\ref{lelem}), one obtains equations of motion for Abelian Higgs vortex in the background in question.
To simplify the relations we redefine our variables by virtue of the following:
\be
\sqrt{\lambda} \eta (r,l) \rightarrow (r,l).
\ee 
Then, equations of motion for the Abelian Higgs vortex on the background in question imply
\ben
A^{2}~{d^2 \over dr^{2}}
X + 
\frac{ 1 }{ \sqrt{-g} }  {d \over dr}~\bigg( \sqrt{-g} A^{2} \bigg) {d \over dr} X -
C^{-2} N^2 P^2 X - 
\frac{ 1 }{ 2 } X \bigg( X^2 - 1 \bigg) = 0, \\
C^{-2} A^{2}{d^2 \over dr^{2}} P +
\frac{ 1 }{ \sqrt{-g} } {d \over dr}~\bigg( \sqrt{-g} A^{2} C^{-2} \bigg) {d \over dr} P -
\frac { 1 }{ \nu } C^{-2} P X^2 = 0.
\een
where we have denoted $\nu = \frac{ \lambda }{ 2 e_{R}^2 }$.\\
The exact solution of the above equations were elaborated in \cite{deh02v}, where as the background metric (\ref{lelem})
the charged black string line element was taken. The metric describing a static charged black string with an Abelian Higgs vortex was
found. It was revealed that the presence of the Higgs fields induced a deficit angle in the charged black string line element.

\section{Fermions in the Black String Spacetime}
In this section we shall pay attention to the fermion superconductivity of a cosmic string which pierces
the charged black string. It was shown \cite{wit85} that in various field theories cosmic strings behave like 
superconducting carrying electric currents. In principle one can distinguished two kinds of this phenomenon. Namely,
we have to do with bosonic or fermionic superconductivity. If a charged Higgs field acquires an expectation value
in the core of the cosmic string we can regard it as bosonic superconductivity. On the other hand,
when Jackiw-Rossi \cite{jac81} charged zero modes appear which can be regarded as Nambu-Goldstone bosons in $1+1$-dimensions,
we obtain fermionic superconductivity. Charged zero modes give rise to a longitudinal components of 
the photon field on the considered cosmic string and may be trapped as massless zero modes.
\par
In order to obtain fermionic superconductivity we extend our $U(1) \times U(1)$ Lagrangian by adding the following one,
for the fermionic sector:
\be
S_{FE} = \int \sqrt{- g}~ d^4x \left [ 
i\bar{\psi}\gamma^{\mu}D_{\mu}\psi +i\bar{\chi}\gamma^{\mu}D_{\mu}\chi 
+ i~ \talpha~ \bigg( \Phi~ \psi^{T} C \chi - \Phi^{*}~ \bar{\psi}C \bar{\chi}^{T} \bigg) \right ],
\ee
where $\talpha$ is a coupling constant. Dirac operator in the above relation yields
\be
D_{\mu} = \nabla_{\mu} + i~ \hat{r}~e_{R}B_{\mu} + i~\hat{q}~e_{q}A_{\mu},
\label{cov}
\ee
where we take as a component of the gauge field  $A_{\mu} = - \frac{l \tilde{\lambda}}{r} \delta^{t}_{\mu} $. 
$\na_{\mu}$ stands for the {\it covariant derivative}
for spinor fields. We choose Dirac gamma matrices which form the chiral basis for the problem in question.
They are of the form as
\be
\gamma^{0} = \pmatrix{ 0&I \cr I&0 } , \qquad
\gamma^{a} = \pmatrix{ 0&\sigma^{a} \cr -\sigma^{a}&0 },
\label{gam}
\ee
where the Pauli matrices are given by
\ben
\sigma^{0} = \pmatrix{ 1&0 \cr 0&1 }, \qquad
\sigma^{1} = \pmatrix{ 0&1 \cr 1&0 }, \\
\sigma^{2} = \pmatrix{ 0&-i \cr i&0 } , \qquad
\sigma^{3} = \pmatrix{ 1&0 \cr 0&-1 }. 
\een
The charge conjugation matrix implies
\ben
C &=& \pmatrix{ -i\sigma^{2}&0 \cr 0&i\sigma^{2} }, \\
C^{\dagger} &=& C^{T} = -C.
\een
The gamma matrices in the curved cylindrically symmetric spacetime under consideration will be provided by
the relations
\ben
\gamma^{t} &=& A^{-1} \pmatrix{ 0&I \cr I&0 } , \qquad
\gamma^{r} =B^{-1} \pmatrix{ 0&\sigma^{1} \cr -\sigma^{1}&0 }, \\
\gamma^{\phi} &=& C^{-1} \pmatrix{ 0& \sigma^{2} \cr -\sigma^{2}&0 } , \qquad
\gamma^{z} =D^{-1} \pmatrix{ 0&\sigma^{3} \cr -\sigma^{3}&0 }.
\een
Spinors $\psi$ and $\chi$ and their Hermitian conjugates should be regarded as the independent fields and the 
variation of the action, that govern them, will lead us to the equations of motion provided by
\ben
\Dsl \psi - \tilde{\alpha} \Phi^{*} C \bar{\chi}^{T} = 0, \\ \nonumber
\Dsl^{\dagger} \bar{\chi}^{\dagger} - \tilde{\alpha} \Phi^{*} C^{\dagger} \psi^{*} = 0. 
\een
The analogous relations will be obtained for their conjugations, while the
Dirac operator may be cast in the form as
\be
\Dsl = \gamma^{\mu}D_{\mu} = 
\pmatrix{ 0& D^{+} \cr D^{-}&0},
\ee
where we have denoted by $D^{+}$ and $D^{-}$ the following parts of the Dirac operator defined above
\ben
D^{+} &=& \sigma^{t}D_{t} + \sigma^{k}~D_{k}, \\
D^{-} &=&  \sigma^{t}D_{t} - \sigma^{j}~D_{j}.
\een
Inserting to these equations the following form of spinors
\be
\psi = \pmatrix {\psi_{L} \cr \psi_{R}}, \qquad
\chi = \pmatrix { \chi_{L} \cr \chi_{R}},  
\ee   
enables us to conclude that chiralities decouple.
\par
Thus, from this stage on we shall consider equations of motion provided by the following relations:
\ben
D^{-} \psi_{L} &-& i \tilde{\alpha} \Phi^{*} \sigma^{2} \chi_{L}^{*} = 0, \\ \nonumber
D^{-} \chi_{L} &-& i \tilde{\alpha} \Phi^{*} \sigma^{2} \psi_{L}^{*} = 0.
\een 
Taking into account that
spinors $\psi_{L}$ and $\chi_{L}$ may be written in the form
\be
\psi_{L} = \pmatrix {f_{+} \cr f_{-}}, \qquad \chi_{L} = \pmatrix {g_{+} \cr g_{-}}, 
\ee   
the underlying equations reduce to the following system:
\ben
\label{ss1}
A^{-1}[\partial_{t} &+& i \hat{q}A_{t}]f_{+} -
\bigg [ B^{-1}\partial_{r}   + \frac{1}{2} A^{-1} B^{-1} \partial_{r}A 
        + \frac{1}{2} B^{-1} C^{-1} \partial_{r} C 
        + \frac{1}{2} B^{-1} D^{-1} \partial_{r} D \bigg]f_{-} \\ \nonumber 
&+& i C^{-1} [\partial_{\phi} + i \hat{r} P_{\phi}]f_{-}  - D^{-1}\partial_{z}f_{+}
- \tilde{\alpha} \Phi^{*} g_{-}^{*} = 0, \\
A^{-1}[\partial_{t} &+& i \hat{q}A_{t}]f_{-} -
\bigg [ B^{-1}\partial_{r}   + \frac{1}{2} A^{-1} B^{-1} \partial_{r}A 
        + \frac{1}{2} B^{-1} C^{-1} \partial_{r} C 
        + \frac{1}{2} B^{-1} D^{-1} \partial_{r} D \bigg]f_{+} \\ \nonumber 
&-& i C^{-1} [\partial_{\phi} + i \hat{r} P_{\phi}]f_{+} + D^{-1}\partial_{z}f_{-}
+ \tilde{\alpha} \Phi^{*} g_{+}^{*} = 0, \\
A^{-1}[\partial_{t} &+& i \hat{q}A_{t}]g_{+} -
\bigg [ B^{-1}\partial_{r}   + \frac{1}{2} A^{-1} B^{-1} \partial_{r}A 
        + \frac{1}{2} B^{-1} C^{-1} \partial_{r} C 
        + \frac{1}{2} B^{-1} D^{-1} \partial_{r} D \bigg]g_{-} \\ \nonumber 
&+& i C^{-1} [\partial_{\phi} + i \hat{r} P_{\phi}]g_{-} - D^{-1}\partial_{z}g_{+} 
- \tilde{\alpha} \Phi^{*} f_{-}^{*} = 0, \\ \label{ss2}
A^{-1}[\partial_{t} &+& i \hat{q}A_{t}]g_{-} -
\bigg [ B^{-1}\partial_{r}   + \frac{1}{2} A^{-1} B^{-1} \partial_{r}A 
        + \frac{1}{2} B^{-1} C^{-1} \partial_{r} C 
        + \frac{1}{2} B^{-1} D^{-1} \partial_{r} D \bigg]g_{+} \\ \nonumber 
&-& i C^{-1} [\partial_{\phi} + i \hat{r} P_{\phi}]g_{+} + D^{-1}\partial_{z}g_{-} 
+ \tilde{\alpha} \Phi^{*} f_{+}^{*} = 0,
\een
where $P_{\phi}$ is the $\phi$-component of $P_\mu$ given by the relation (\ref{phi}).\\
Let us suppose that the explicit forms of the functions $f$ and $g$ imply
\ben
f_{\pm} &=& e^{\epsilon i(\omega t - kz)}e^{im_{1/2}\phi}f_{\pm}(r), \\
g_{\pm} &=& e^{\epsilon i(\omega t - kz)}e^{im_{3/4}\phi}g_{\pm}(r),  
\een
where $\epsilon$ is equal to $\pm 1$.\\
As was pointed out in Ref.\cite{wit85} the role of fermions became interesting if we considered the fermions which 
gained their masses from coupling to $\Phi$ field. Just, the action of the operators $\hat{r}$ and $\hat{q}$
in the definition of the Dirac operator (\ref{cov}) yield
\ben
\hat{q}~e_{q}~f &=& q_{e}~ f, \qquad \hat{q}~e_{q}~g = - q_{e}~ g, \\
\hat{r}~f &=& q_{r}~ f , \qquad \hat{r}~g = - (q_{r} + 1)~ g,
\een
where $q_{e}$ and $q_{r}$ are spinor charges. Due to the above relations, we obtained that
fermion fields $\psi$ and $\chi$ gained masses from their coupling to $\Phi$-field.
Consequently, having in mind the above relations and the angular dependence of
$\Phi$ field, we conclude that the following ought to be satisfied:
\ben
m_{1} &=& m_{2} = - N - m_{4}, \\
m_{2} &=& m_{1} = - N - m_{3}, \\
m_{3} &=& m_{4} = - N - m_{2}, \\
m_{4} &=& m_{3} = - N - m_{1}. 
\een 
We can readily choose the following angular dependence for fermion fields
\ben
m_{1} &=& m_{2} \equiv m, \\
m_{4} &=& m_{3} = -N - m.
\een
By virtue of the above relations, one arrives at
\ben
f_{\pm} &=& e^{ \epsilon i(\omega t - kz) } e^{ im \phi } f_{\pm}(r), \\
g_{\pm} &=& e^{ \epsilon i(\omega t - kz) } e^{-i(N+m) \phi } g_{\pm}(r).  
\een
On the other hand, we want fermions to propagate along the cosmic string so one 
chooses the spinors in question as the eigenstates of $\gamma^0 \gamma^3$ matrices. Namely, we have the following:	
\ben
\gamma^0 \gamma^3 \psi &=&  \psi, \\
\gamma^0 \gamma^3 \chi &=&  \chi.
\een
Having in mind that the chiralities decouple, the spinor functions under consideration imply
\be
\psi_{L} = \pmatrix { 0 \cr f_{-}}, \qquad
\chi_{L} = \pmatrix { 0 \cr g_{-}}. 
\label{exactf}
\ee   
It can be seen that starting with the exact form of the spinors given by (\ref{exactf})
we arrive at the following forms of equations (\ref{ss1})-(\ref{ss2}):
\ben \label{pop1}
- \bigg [ B^{-1}\partial_{r}   + \frac{1}{2} A^{-1} B^{-1} \partial_{r}A 
        + \frac{1}{2} B^{-1} C^{-1} \partial_{r} C 
        + \frac{1}{2} B^{-1} D^{-1} \partial_{r} D \bigg]f_{-}  
+ i C^{-1} [\partial_{\phi} + i \hat{r} P_{\phi}]f_{-}  
- \tilde{\alpha} \Phi^{*} g_{-}^{*} = 0, \\ \label{pop2}
A^{-1}[\partial_{t} + i \hat{q}A_{t}]f_{-}
+ D^{-1}\partial_{z}f_{-} = 0, \\ \label{pop3}
- \bigg [ B^{-1}\partial_{r}   + \frac{1}{2} A^{-1} B^{-1} \partial_{r}A 
        + \frac{1}{2} B^{-1} C^{-1} \partial_{r} C 
        + \frac{1}{2} B^{-1} D^{-1} \partial_{r} D \bigg]g_{-} 
+ i C^{-1} [\partial_{\phi} + i \hat{r} P_{\phi}]g_{-} 
- \tilde{\alpha} \Phi^{*} f_{-}^{*} = 0, \\ \label{pop4}
A^{-1}[\partial_{t} + i \hat{q}A_{t}]g_{-} 
+ D^{-1}\partial_{z}g_{-} = 0.
\een
To proceed further, let us choose the following form of $f_{-}$ and $g_{-}$ spinors:
\be
f_{-} = e^{\epsilon i(\omega t - k z)}
e^{i m \phi}
e^{- \int \bigg[ \frac{1}{2} B^{-1} C^{-1} \partial_{r} C +  
 \frac{1}{2} B^{-1} D^{-1} \partial_{r} D \bigg] B dr } \tilde{f}_{-}, 
 \label{de1}   
\ee
\be
g_{-} = e^{\epsilon i(\omega t - k z)}
e^{- i (N + m) \phi}
e^{- \int \bigg [ \frac{1}{2} B^{-1} C^{-1} \partial_{r} C + 
                  \frac{1}{2} B^{-1} D^{-1} \partial_{r} D \bigg] B dr } \tilde{g}_{-}.
\label{de2}
\ee
The explicit forms of the metric coefficients $B,~C,~D$ envisage the fact that the above integrals
are convergent in the limit of $r$-coordinate tends to infinity. By virtue of the numerical
integration, one can check that this conclusion is also true for the finite $r$.\\
From equations (\ref{pop1}) and (\ref{pop3}) we obtain 
\ben
\label{system-g0g3}
\bigg [ B^{ -1 } \partial_{r} 
&+& \frac{ 1 }{ 2 } A^{-1} B^{-1} \partial_{r} A 
+ C^{ -1 } \bigg( m + q_{r} P_{\phi} \bigg) \bigg ] \tilde{f}_{-} +  
m_{fer} X \tilde{g}_{-} = 0, \\ \label{system-g0g31}
\bigg [ B^{-1} \partial_{r} 
&+& \frac{ 1 }{ 2 } A^{-1} B^{-1} \partial_{r} A 
- C^{ -1 } 
\bigg( N + m + ( q_{r} + 1 ) P_{\phi} \bigg) \bigg ] \tilde{g}_{-} + 
m_{fer} X \tilde{f}_{-} = 0,                     
\een
where we have denoted $m_{fer} = \tilde{\alpha} \eta$.
On the other hand, the second and the fourth relations of the system in question give us
\ben
- i[ A^{-1} ( \omega + q_{e} \frac{ l \tilde{\lambda} }{ r }  ) - \frac{ k }{ D } ]
e^{ - i (\omega t - k z ) }e^{ i m \phi}
e^{- \int \bigg[ \frac{1}{2} B^{-1} C^{-1} \partial_{r} C +  
 \frac{1}{2} B^{-1} D^{-1} \partial_{r} D \bigg] B dr } \tilde{f}_{-}(r) = 0, \\ 
i[ A^{-1}(\omega + q_{e} \frac{ l \tilde{\lambda} }{ r } ) - \frac{ k }{ D } ]
e^{  i (\omega t - k z ) }e^{ - i ( N+ m ) \phi}
e^{- \int \bigg[ \frac{1}{2} B^{-1} C^{-1} \partial_{r} C +  
 \frac{1}{2} B^{-1} D^{-1} \partial_{r} D \bigg] B dr } \tilde{g}_{-}(r) = 0.
\een
Just, from the inspection of the above equations we have either $\tilde{f}_{-} = \tilde{g}_{-} = 0$ or
\ben
[ A^{-1} ( \omega + q_{e} \frac{ l \tilde{\lambda} }{ r }  ) - \frac{ k }{ D } ] = 0.
\een
Because of the fact that we have used the ansatze (\ref{de1}) and (\ref{de2}), one has that
that $\omega, k , q_{e}$ are constant and equal to zero. Therefore,  
the relation $\omega = k = q_{e} = 0$ emerges as the {\it consistency} condition for the system
of equations (\ref{system-g0g3}) - (\ref{system-g0g31}). It can be remarked that the procedure
as described above leads to {\it normalizable fermionic zero modes}.
\par
From now on,
for the brevity of the subsequent notation, we define the rescaled version of the parameters
characterizing fermion fields. Namely, we set the following:
\be
\frac { 1 }{ \sqrt{\lambda} \eta }( \omega, q_{e}, k, m_{f} ) \equiv  ( \omega, q_{e}, k ,m_{f} ). 
\ee

\section{Asymptotic behaviour of spinor fields in the background of a charged cosmic string}
In this section we treat first the behaviour of fermions in the distant region from the considered
charged black string pierced by an Abelian Higgs vortex. 
In order to simplify our notation we redefine once more the spinor functions in question, by the relations
\be
\tilde{f}_{-} = \frac{ 1 }{ \sqrt{A} } \tilde{f}_{-}, \qquad
\tilde{g}_{-} = \frac{ 1 }{ \sqrt{A} } \tilde{g}_{-}.
\ee
On this account, one can rewrite the underlying equations of motion as follows:
\ben
\bigg [ B^{ -1 } \partial_{r} 
&+& C^{ -1 } [ m + q_{r} (P - N ) ] \bigg ] f_{-} + 
m_{fer} X g_{-} = 0, \\
\bigg [ B^{-1} \partial_{r} 
&-& C^{ -1 } [ N + m + ( q_{r} + 1 ) ( P - N) ] \bigg ] g_{-} + 
m_{fer} X f_{-} = 0.
\een
Asymptotically, when $r \rightarrow \infty$ the value of $P$ field tends to zero, while $X = 1$.
Hence, our equations reduce to the system of differential equations given by
\ben
\bigg [ B^{ -1 } \partial_{r} 
&+& C^{ -1 } [ m  - q_{r} N ] \bigg ] \tilde{f}_{-} + 
m_{fer} \tilde{g}_{-} = 0, \\
\bigg [ B^{-1} \partial_{r} 
&-& C^{ -1 } [ m - q_{r}N ] \bigg ] \tilde{g}_{-} + 
m_{fer} \tilde{f}_{-} = 0,
\een
which can be brought to the second order differential equation provided by
\ben
\tilde{g}_{-} &=& -  \frac { 1 }{m_{fer}}
\bigg [ \partial_{r^{*}} \tilde{f}_{-} + 
 C^{-1} [ m - q_{r} N] \tilde{f}_{-} \bigg ], \\ \nonumber
\partial_{r^{*}}^2 \tilde{f}_{-} &-& 
\bigg [  m_{fer}^2 + 
\frac  { (m - q_{r} N )^2 } { C^2 } -
\frac  { m - q_{r} N } { C^2 } \frac { d C } { d r} \frac{ d r } { d r^{*} } 
\bigg ] \tilde{f}_{-} = 0,
\een
where $B^{-1} \partial_{r} \equiv \partial_{r^{*}}$.\\
In order to estimate the asymptotical value of the fermion functions given by the above equations we use the theorem 
\cite{Hartman}, which states that for the second order differential equation of the type 
$\frac{d^2}{d r^2} u - [c^2 + q(r)]u=0$, there exists an asymptotic solution in the form 
$u_{\pm} \sim  c_{\pm} e^{\pm c r}$ if $\int^{\infty} |q(r)|dr < \infty$.  
It may be noted that in our case $|q(r)|$ implies the following:
\be
|q| = q(r^{*}) = \frac  { (m - q_{r} N )^2 } { C^2 } -
\frac  { m - q_{r} N } { C^2 } \frac { d C } { d r} \frac{ d r } { d r^{*} }.
\ee
Further on, carrying out the integration of the above relation we arrive at
\be
\int^{\infty} q( r^{*}) d r^{*} = 
\int^{\infty} \frac  { (m - q_{r} N )^2 } { C^2 } d r^{*} -
\int^{\infty} \frac  { m - q_{r} N } { C^2 } \frac { d C } { d r} \frac{ d r } { d r^{*} } d r^{*}.
\ee
The second integral on the right-hand side is straightforward to perform and it yields $\frac{ m - q_{r} N} { C }$.
As far as the first one is concerned, it yields
\be
\int^{\infty} \frac  { (m - q_{r} N )^2 } { C^2 } d r^{*}  = 
(m - q_{r} N )^2~  l~ \int^{\infty} \frac {   d r} 
{ \sqrt { r^6 - b l^3 r^3 + \tilde{\lambda}^2 l^4 r^2 }}.
\ee 
The above integral can be easily calculated numerically. It happens that it has a finite value 
when $r$ tends to infinity. Thus having in mind the quoted theorem, the asymptotical solutions for 
$\tilde{f}$ and $\tilde{g}$ functions are of the form $c_{\pm}e^{\pm m_{fer}r^{*}}$.
In order to situate fermions inside the cosmic string we set $c_{+} = 0$. By virtue of this requirement we get

\ben
\label{solution-infty-g0g3}
\tilde{f}_{-}(r^{*} \rightarrow \infty) &=& \frac{ c_{-} } { \sqrt{ A } } e^{- m_{fer} r^{*}},\\
\tilde{g}_{-}(r^{*} \rightarrow \infty) &=& \frac{ c_{-} } { \sqrt{ A } } e^{- m_{fer} r^{*}}  
\bigg [ 1 - \frac {  m - q_{r} N }{ m_{fer} C } \bigg ].
\een

\section{Nonextremal charged black string and fermion fields}
\subsection{Electrically uncharged fermions. }
In order to study the behaviour of fermion fields in the vicinity of black sting event horizon we expand
the metric coefficients in the nearby of black string horizon in the forms as follows:
\ben
A^{2}(r_{h}) \sim  \partial_{r}A^{2}_{r = r_{h}} (r - r_{h}) &=& 2 \kappa (r - r_{h}),\\
C^2(r_{h}) &=& r_{h}^2,
\een
where the surface gravity $\kappa$ is given by standard formula $\kappa = \frac{ 1 }{ 2 } \partial_{r} A^2 _{| r_{h}}$.
By virtue of the above relations the line element describing the near-horizon black string geometry implies
\be
ds^2 = - 2 \kappa\rho^2 dt^2 + \frac{ 2 }{ \kappa }d\rho^2 + r_{h}^2 d\phi^2 + \frac { r_{h}^2 } { l^2 }dz^2.
\ee
Introducing new variables and having in mind behaviour of P and X field near horizon we get 
\ben
\rho^2 = r - r_{h}, \qquad
X(\rho \sim 0) = \rho^N, \qquad P(\rho \sim 0) = N + {\cal O}(\rho^2).
\een 
In this picture, the equations of motion will be given by
\ben \label{a1}
\bigg [ \partial_{\rho}  &+& \frac { 1 } { 2 \rho } +  \frac{\sqrt{2} m }{ r_{h} \sqrt{ \kappa } } \bigg ] \tilde{f}_{-} +
\frac { \sqrt{2} m_{fer}} { \sqrt{\kappa} } \rho^N \tilde{g}_{-} = 0, \\ \label{a2}
\bigg [ \partial_{\rho}  &+& \frac { 1 } { 2 \rho } - \sqrt{2} \frac{ N + m }{ r_{h}  \sqrt{ \kappa } } \bigg ] \tilde{g}_{-} +
\frac { \sqrt{2} m_{fer}} { \sqrt{ \kappa } } \rho^N \tilde{f}_{-} = 0. 
\een
Consider now the case when $ N  > > 1$. The mass term is proportional to
$\rho^N \rightarrow 0$ in this case. Then, the solutions of (\ref{a1}) and (\ref{a2}) are provided by
\ben
\tilde{f}_{-} &=& c_{1}~ \rho^{- \frac{ 1 } { 2 } } 
e^{- \frac { \sqrt{2} m } { r_{h} \sqrt{ \kappa } } \rho}, \\
\tilde{g}_{-} &=& c_{2}~ \rho^{- \frac{ 1 } { 2 } } 
e^{ \sqrt{ 2 } \frac { N + m } { r_{h} \sqrt{ \kappa } } \rho}. 
\een 
Consequently, one can readily verify that although these solutions are divergent at the 
black string event horizon, they are square integrable. Namely, they satisfy
\ben
\int_{ 0 }^{ \rho_{max}} \sqrt {-g} |\tilde{f}_{-}|^2 d \rho &=&
\int_{ 0 }^{ \rho_{max}} 2 \frac{ r_{h}^2 }{ l }  \rho ~ 
|c_{1}|^2 \rho^{-1} e^{- 2 \frac { \sqrt{2} m } { r_{h} \sqrt{\kappa} } \rho} d \rho 
 < \infty, \\
\int_{ 0 }^{ \rho_{max}} \sqrt {-g} |\tilde{g}_{-}|^2 d \rho &=&
\int_{ 0 }^{ \rho_{max}} 2 \frac{ r_{h}^2 }{ l }  \rho ~ 
|c_{2}|^2 \rho^{-1} e^{ 2 \sqrt{2} \frac { N + m } { r_{h} \sqrt{ \kappa} } \rho} d \rho 
 < \infty.
\een  
Next, we proceed to the case $ N \sim 1$. It turned out that the above Eqs. for the case in question can be simplified
when we substitute
\be
\tilde{f}_{-} = \tilde{f}_{-}(\bar{m}_{fer} = 0) \bar{f}, \qquad
\tilde{g}_{-} = \tilde{g}_{-}(\bar{m}_{fer} = 0) \bar{g},
\ee
where $\bar{m}_{fer} \equiv \frac{ \sqrt{ 2 } m_{fer}}{ \sqrt{ \kappa }}$.
We denote by 
$\tilde{f}_{-} ( \bar{m}_{fer} = 0 )$ and $\tilde{g}_{-} ( \bar{m}_{fer} = 0 )$ the solutions
of the equations of motion for the case $ N > > 1$.
On this account, the underlying relations imply
\ben
\partial_{\rho} \bar{f} &+& 
\bar{m}_{fer} \rho^N e^{ \sqrt{2} \frac{ N + 2m }{ r_{h} \sqrt{ \kappa } } \rho} \bar{g} = 0, \\
\partial_{\rho} \bar{g} &+& 
\bar{m}_{fer} \rho^N e^{ - \sqrt{2} \frac{ N + 2m }{ r_{h} \sqrt{ \kappa } } \rho} \bar{f} = 0. 
\een
Extracting $\bar{g}$ from the first equation and substituting to the second one, the considered
system of the first order differential equations can be brought to the second order differential equation for $\bar{f}$.
It yields
\ben
\label{near-horizon-N1}
\bar{g} &=& - \frac{ 1 }{ \bar{m}_{fer} }
\rho^{-N} e^{ - \sqrt{2} \frac{ N + 2m }{ r_{h} \sqrt{ \kappa } } \rho} \partial_{\rho} \bar{f}, \\
\partial_{\rho}^2 \bar{f} &-&
\bigg [ N \rho^{-1} + \bar{b} \bigg] \partial_{\rho} \bar{f} -
\bar{m}_{fer}^2 \rho^{ 2N } \bar{f} = 0,
\een
where we put $ \bar{b} = \sqrt{2} \frac{ N + 2m }{ r_{h} \sqrt{ \kappa } }$.
Unfortunately, these equations have no solutions in terms of the known special functions which implies that
they should be treated numerically.

\subsection{Electrically  charged fermions.}
The next object of an interest is the influence of electrically charged fermions on the superconductivity of the cosmic
string which pierced the charged black string.
In order to find the simplest electrically
charged solution of Eqs.(\ref{ss1})-(\ref{ss2})
we use the linear combination of 
spinors being the eigenstates  of $\gamma^0 \gamma^3$. Under this assumption
we take into account spinor fields $\psi_{L}$ and $\chi_{L}$ provided by the relations
\be
\psi_{L} = \pmatrix{ if_{-} \cr f_{-}}, \qquad
\chi_{L} = \pmatrix{ - ig_{-} \cr g_{-}}.
\ee  
Next we use the fact that fermion functions depend only on 
($t,z,\phi$)-coordinates. Namely, we have
\ben
f_{-} = e^{ - i ( \omega t - kz ) }e^{ i m \phi } \bar{f}_{-}, \\
g_{-} = e^{  i ( \omega t - kz ) }e^{ - i ( N + m ) \phi } \bar{g}_{-}. 
\een
Now, the system of equations (\ref{ss1})-(\ref{ss2}) reduces to the relations given by
\ben
\label{system-full}
A^{-1}[\omega &+& q_{e} \frac{ \tilde{\lambda } l }{ r } ] \bar{f}_{-} -
\bigg [ B^{-1}\partial_{r}   + \frac{1}{2} A^{-1} B^{-1} \partial_{r}A 
        + \frac{1}{2} B^{-1} C^{-1} \partial_{r} C 
        + \frac{1}{2} B^{-1} D^{-1} \partial_{r} D \bigg] \bar{f}_{-} \\ \nonumber 
&-& C^{-1} [m + q_{r} ( P - N) ] \bar{f}_{-}  + k D^{-1} \bar{f}_{-}
+ m_{fer} X \bar{g}_{-} = 0, \\
A^{-1}[\omega &+& q_{e} \frac{ \tilde{\lambda } l }{ r }  ] \bar{g}_{-} -
\bigg [ B^{-1}\partial_{r}   + \frac{1}{2} A^{-1} B^{-1} \partial_{r}A 
        + \frac{1}{2} B^{-1} C^{-1} \partial_{r} C 
        + \frac{1}{2} B^{-1} D^{-1} \partial_{r} D \bigg] \bar{g}_{-} \\ \nonumber 
&+&  C^{-1} [ N + m + (q_{r} + 1 ) ( P - N) ] \bar{g}_{-} + k  D^{-1}\bar{g}_{-} 
- m_{fer} X \bar{f}_{-} = 0.
\een
As in the uncharged fermion case we can decompose functions $f_{-}$ and $g_{-}$ in the following way:
\be
f_{-} = e^{- i(\omega t - k z)}
e^{i m \phi}
e^{- \int \bigg [ \frac{1}{2} B^{-1} C^{-1} \partial_{r} C +  
                  \frac{1}{2} B^{-1} D^{-1} \partial_{r} D 
                  - \frac{ k }{ D } - A^{-1} ( \omega - q_{e} A_{t})  \bigg ] B dr } \tilde{f}_{-}, 
\label{cde1}
\ee
\be
g_{-} = e^{ i(\omega t - k z)} e^{- i (N + m) \phi}
e^{- \int \bigg [ \frac{1}{2} B^{-1} C^{-1} \partial_{r} C + 
                  \frac{1}{2} B^{-1} D^{-1} \partial_{r} D 
                  - \frac{ k }{ D } - A^{-1} ( \omega - q_{e} A_{t})  \bigg] B dr } \tilde{g}_{-}, 
\label{cde2}
\ee
where $A,~B,C,~D$ are the functions from the line element describing charged black string.
Having in mind the explicit forms of them, one can see that the above integrals are convergent in the limit
of $r \rightarrow \infty$. On the other hand, the explicit use of numerical integrations confirms this fact
for the finite value of $r$-coordinate.\\
On this account, it is customary to write the system of equations (\ref{ss1})-(\ref{ss2}) in the form as
\ben
\label{system-charged}
B^{-1}\partial_{r} \tilde{f}_{-} &+& 
\bigg[\frac{ 1 }{ 2 } B^{-1} A^{-1} \partial_{r} A + 
\frac{ m + q_{r}(P - N) }{ C } \bigg]~\tilde{f}_{-}  + m_{fer} X \tilde{g}_{-} = 0, \\
B^{-1} \partial_{r} \tilde{g}_{-} &+& 
\bigg[ \frac{ 1 }{ 2 } B^{-1} A^{-1} \partial_{r} A - 
\frac{ m + N + (q_{r} + 1)(P - N) }{ C } \bigg]~\tilde{g}_{-} + m_{fer} X \tilde{f}_{-} = 0. 
\een
Thus the asymptotic form of the functions in question may be written as
\ben
\tilde{f}_{-}( r^{*} \rightarrow \infty ) &=& \frac{ c_{-} } { \sqrt{ A } } e^{- m_{fer} r^{*}}, \\
\tilde{g}_{-}( r^{*} \rightarrow \infty ) &=& \frac{ c_{-} } { \sqrt{ A } } e^{ - m_{fer} r^{*}}
\bigg[1 - \frac{ m - q_{r}N }{ m_{fer} C } \bigg].
\een
On the other hand, in the near-horizon limit we get the following system of equations:
\ben
\label{system-nearh-charged}
\partial_{\rho} \tilde{f}_{-} &+& 
\bigg[ \rho^{-1} \bigg( \frac{ 1 }{ 2 } - \frac{ q_{e} \tilde{\lambda} l}{ \kappa  r_{h} } 
\bigg) \bigg]~\tilde{f}_{-} + 
\frac{ \sqrt{2} m }{ \sqrt{ \kappa } r_{h}  }\tilde{f}_{-}  + \bar{m} \rho^{N} \tilde{g}_{-} = 0, \nonumber \\
\partial_{\rho} \tilde{g}_{-} &+& 
\bigg[ \rho^{-1} \bigg( \frac{ 1 }{ 2 } - \frac{ q_{e} \tilde{\lambda} l}{  \kappa  r_{h} } \bigg)
\bigg]~\tilde{g}_{-} - 
 \sqrt{2} \frac{ N + m }{ \sqrt{ \kappa} r_{h}  }\tilde{g}_{-}  + \bar{m} \rho^{N} \tilde{f}_{-} = 0. 
\een
where we put for the brevity $\bar{m} = \frac{ \sqrt{2} m_{fer} }{ \sqrt{ \kappa } }$.\\
By virtue of the above we conclude that for $N >> 1$ fermions in the nearby of the black string event 
horizon are essentially massless. Explicitly, they read
\ben
\tilde{f}_{-} &=& c_1~ \rho^{{ q_e~\tilde{\lambda}~l \over \kappa~r_h}  - \frac{ 1 }{ 2 } }
                e^{ - \sqrt{ 2 } \frac{ m }{ \sqrt{\kappa} r_{h} }  \rho }, \\
\tilde{g}_{-} &=& c_2~ \rho^{ \frac{  q_{e} \tilde{\lambda} l}{  \kappa  r_{h} }   - \frac{ 1 }{ 2 } }
                e^{  \sqrt{ 2 } \frac{ N + m }{ \sqrt{\kappa} r_{h} } \rho }.
\een 
One can easily find by the direct calculation that they are square integrable.
On the other hand, for the case when the winding number $ N \sim 1$, we make the following substitution :
\ben
\tilde{f}_{-} = \tilde{f}_{-}(\bar{m}_{fer} = 0) \bar{f}, \qquad
\tilde{g}_{-} = \tilde{g}_{-}(\bar{m}_{fer} = 0) \bar{g},
\een
where $\tilde{f}_{-} ( \bar{m}_{fer} = 0 )$ and $\tilde{g}_{-} ( \bar{m}_{fer} = 0 )$ are the solutions for $ N > > 1$ case.
Making use of the above substitutions one can readily see that we arrive at the following:
\ben
\partial_{\rho} \bar{f} &+& 
\bar{m}_{fer} \rho^N e^{ \sqrt{2} \frac{ N + 2m }{ r_{h} \sqrt{ \kappa } } \rho} \bar{g} = 0, \\
\partial_{\rho} \bar{g} &+& 
\bar{m}_{fer} \rho^N e^{ - \sqrt{2} \frac{ N + 2m }{ r_{h} \sqrt{ \kappa } } \rho} \bar{f} = 0.
\een
Of course, we can extract $ \bar{g}$ function from the first equation and obtain the second order differential equation for
$\bar{f}$, but it has no solutions in terms of the known special functions.

\section{Extremal black string and fermion fields}
\subsection{Electrically uncharged fermions}
In what follows we shall elaborate some main features of the behaviour of fermion fields in the vicinity of an 
extremal charged black string.
We expand coefficient of the metric in the form as follows:
\be
A^{2}(r) \sim \frac{ 1 }{ 2 } \partial_{r}^{2} A^{2}_{ |r = r_{h} } (r - r_{h})^2.
\ee
It enables us to rewrite the line element in the near-horizon limit in the form
\be
ds^2 = - a(r_{h}) \rho^2 dt^2 + \frac{ d \rho^2 }{ a(r_{h}) \rho^2 } + r_{h}^2 d \phi^2 + \frac{ r_{h}^2 }{ l^2 } dz^2, 
\ee
where $a(r_{h})  = \frac{ 1 }{ 2 } \partial_{r}^{2} A^{2}_{| r = r_{h}} $ and $ \rho \equiv r - r_{h} $.
The near horizon behaviour of the Abelian Higgs vortex fields $P$ and $X$ are given by
\be
X (\rho \rightarrow 0) \sim \rho^{ \frac{ |N| }{ 2 } }, \qquad
P( \rho \rightarrow 0) \sim N + {\cal O}( \rho ).
\ee
Returning to the equation of motion for the uncharged fermions,
one can readily verify that they reduces to the forms
\ben
\sqrt{ a(r_{h}) } \rho \partial_{\rho} \tilde{f}_{-} &+& 
\bigg[ \frac{ 1 }{ 2 } \sqrt{ a(r_{h}) } + \frac{ m }{ r_{h} } \bigg]~\tilde{f}_{-} + 
m_{fer} \rho^{ \frac{ |N| }{ 2 } } \tilde{g}_{-} = 0, \\
\sqrt{ a(r_{h}) } \rho \partial_{\rho} \tilde{g}_{-} &+&
\bigg[ \frac{ 1 }{ 2 } \sqrt{ a(r_{h}) } - \frac{ N + m }{ r_{h} } \bigg]~\tilde{g}_{-} + 
m_{fer} \rho^{ \frac{ |N| }{ 2 } } \tilde{f}_{-} = 0. 
\een 
First, we shall consider the influence of the winding number $N$ on the behaviour of fermion fields in question.
For sufficiently large N and small $\rho$ one can neglect the mass term and the solutions
are provided by the following relations: 
\ben
\tilde{f}_{-} &=& c_1~ \rho^{ - ( \frac{ 1 }{ 2 } + \frac{ m }{ \sqrt{ a(r_{h} ) r_{h} } } )}, \\
\tilde{g}_{-} &=& c_2~ \rho^{ - ( \frac{ 1 }{ 2 } - \frac{ m + N }{ \sqrt{ a(r_{h}) } r_{h} }  )  }
\een
One can observe that they are divergent as in the nonextremal case. \\
On the other hand, for small $N$ we seek solution in the following form:
\ben
\tilde{f}_{-} &=& \tilde{f}_{-}( m_{fer} = 0 ) \bar{f}, \\
\tilde{g}_{-} &=& \tilde{g}_{-}( m_{fer} = 0 ) \bar{g}. 
\een
On this account, we get the system of the first order differential equations provided by
\ben
\partial_{\rho} \bar{f} &+& \hat{m}_{fer} \rho^{ \frac{ |N| }{ 2 } - 1 + \frac{ N + m }{ \sqrt{ a(r_{h}) } r_{h} } } \bar{g} = 0, \\
\partial_{\rho} \bar{g} &+& \hat{m}_{fer} \rho^{ \frac{ |N| }{ 2 } - 1 - \frac{ N + m }{ \sqrt{ a(r_{h}) } r_{h} } } \bar{f} = 0, 
\een
where we have denoted by $\hat{m}_{fer} = \frac{ m_{fer} }{ \sqrt{ a(r_{h}) } }$.\\
The aforementioned system of differential equations can be rearrange in the form of the single 
second order differential equation by the following transformation:
\ben
\label{nearh-extr}
\bar{g} &=& - \frac{ 1 }{ \hat{m}_{fer} } \rho^{ \bar{b} } \partial_{ \rho} \bar{f}, \\
\partial_{ \rho}^2 \bar{f} &+& \frac{ \bar{b} }{ \rho } \partial_{\rho} \bar{f} -
\hat{ m }_{fer}^2 \rho^{ |N| - 2 } \bar{f} = 0,
\een
where $\bar{b} = 1 - \frac{ |N| }{ 2 } - \frac{ N + 2 m }{ \sqrt{ a(r_{h}) } r_{h} }$.
We shall look for the solution of the above equation making the so-called Lommel's transformation for Bessel functions.
Namely, we shall consider the solution in the form
\be
\bar{f} = \rho^p~G_{\nu} \bigg( \la~\rho^q \bigg),
\ee
where $G_{\nu}$ stands for the adequate Bessel function, while $p,~\la,~q$ denote the constants.
It happened that the solution in question can be provided by the function
\be
\bar{f} = c_1~ \rho^{ \frac{ 1 - \bar{b} }{ 2 } } I_{\nu} \bigg( 2~{i m \over N}~\rho^{N \over 2} \bigg)
+ c_2~ \rho^{ \frac{ 1 - \bar{b} }{ 2 } } K_{\nu} \bigg( 2~{i m \over N}~\rho^{N \over 2} \bigg),
\ee
where $\nu = {1 -  \bar{b} \over N}$. When we choose $c_2 = 0$, then from the asymptotic value of $I_\nu$ function 
the solution tends to the finite value.

\subsection{Electrically charged fermions.}
For electrically charged fermions our equation have the following forms:
\ben
\partial_{\rho} \tilde{f}_{-} &+& 
\bigg[ \bigg( \frac{ 1 }{ 2 } + \frac{ m }{ \sqrt{ a(r_{h} ) } r_{h} } \bigg)~ \rho^{ -1 } - 
  q_{e} \frac{ \tilde{\lambda} l }{ \sqrt{ a(r_{h}) } r_{h} } \rho^{ -2 } \bigg]~\tilde{f}_{-} + 
\hat{m}_{fer} \rho^{ \frac{ |N| }{ 2 } -1 } \tilde{g}_{-} = 0, \\
\partial_{\rho} \tilde{g}_{-} &+& 
\bigg[ \bigg( \frac{ 1 }{ 2 } - \frac{ m + N }{ \sqrt{ a(r_{h} ) } r_{h} } \bigg)~ \rho^{ -1 } - 
  q_{e} \frac{ \tilde{\lambda} l }{ \sqrt{ a(r_{h}) } r_{h} } \rho^{ -2 } \bigg]~\tilde{g}_{-} + 
\hat{m}_{fer} \rho^{ \frac{ |N| }{ 2 } -1 } \tilde{f}_{-} = 0.
\een 
A close inspection reveals that for the case $N \gg 1$ one arrives at the relations given by
\ben
\tilde{f}_{-} &=& c_1~ \rho^{ - ( \frac{ 1 }{ 2 } + \frac{ m }{ \sqrt{ a(r_{h}) r_{h}} } ) }
e^{ - q_{e} \frac{ \tilde{\lambda} l}{ \sqrt{ a(r_{h} ) } r_{h} } \rho^{ - 1} }, \\
\tilde{g}_{-} &=& c_2~ \rho^{ - ( \frac{ 1 }{ 2 } - \frac{ m + N }{ \sqrt{ a(r_{h}) r_{h}} } ) }
e^{ - q_{e} \frac{ \tilde{\lambda} l}{ \sqrt{ a(r_{h} ) } r_{h} } \rho^{ - 1} }. 
\een
We observe that for sufficiently large electric charge
$q_{e}$ the exponential term becomes dominant as $\rho$ tends to zero. Moreover,
the underlying solution becomes finite at the event horizon of the extremal charged black string.
\par
For small value of the winding number $N$ we use the substitution in the form as
\ben
\tilde{f}_{-} &=& \tilde{f}_{-}(\hat{m}_{fer} = 0) \bar{f}, \\
\tilde{g}_{-} &=& \tilde{g}_{-}(\hat{m}_{fer} = 0) \bar{g}, 
\een   
which enables us to rewrite Eqs. of motion in 
the same form of the second order differential equation like in the uncharged case, given by the relation (\ref{nearh-extr}).

\section{Numerical solution.}
In order to solve numerically the system of the differential equations 
describing behaviour of the Dirac fermions in the spacetime of a charged black string, first one ought to
find the solutions of equations of motion for the Higgs fields $X$ and $P$. The boundary
conditions for $X$ and $P$ are chosen in such a way that for the large distances from a charged black string
horizon one achieves the vortex solution in AdS spacetime \cite{deh02b}, which means that $X \rightarrow 1$
and $P \rightarrow 0$ as $r$-coordinate tends to infinity. On the other hand, on the black string horizon we
assume that $X = 0$ and $P = 1$ as was done in Ref.\cite{deh02v}. Then, the {\it relaxation method} was used
to obtain solutions for the Higgs fields in the interval $< r,~r_{max}>$, where as $r_{max}$ we set
$r_{max} = 20~r_h$.
\par
Next, we transform the infinite domain $<r_{h},~\infty )$ of $r$-coordinate to the finite one
using the transformation of  the form $z = \frac{ 1 }{ r }$. We also perform this transformation
in the case of the fermion equations of motion and convert the $r$-dependence of $X$ and $P$ functions
to $z$-dependence. We stretch $X$ and $P$ functions to the whole $z$-domain by adding points in the
interval in question and assigning with them the asymptotic values of the considered Higgs fields $X$ and $P$.
To proceed further, one should have the values of $X$ and $P$ in subintervals of equal length
in $z$-direction. 
It
was accomplished by the {\it cubic spline interpolation method} \cite{press}.\\
The last step was to solve numerically equations of motion for Dirac fermions. To begin with, we use
the analytic form of the fermion functions $f_{-}$ and $g_{-}$ at infinity given by the relation
(\ref{solution-infty-g0g3}) and the formulae (\ref{de1})-(\ref{de2}) in the uncharged fermion case
as well as the relations (\ref{cde1})-(\ref{cde2}) in the charged fermion case.
We start our numerical computations from $z =10^{-5}$.\\
Using the {\it implicit trapezoidal method} \cite{press} we propagate these functions up to the charged black string
event horizon and solve the neutral zero-energy fermion set of equations (\ref{system-g0g3}) and the set of 
relations describing charged fermions (\ref{system-charged}). 
\par
In our considerations, studying the nonextremal black string superconducting cosmic string system
we set that $b = 2b_{crit}$, $\tilde{\lambda} = 0.5$ and $ l = 1.0$. The charged black string event horizon was located 
at  $r_{h} = 0.9966$ (nonextremal black string) and $r_{h} = 0.5373$ (extremal black string).
Moreover, the fermion fields
in question will satisfy the normalization condition provided by 
\be
\int_{r_{h}}^{\infty} \sqrt{ -g }~ \xi_{i}^{\dagger} \xi^{i} dr = 1,
\ee 
where by $\xi_{i}$ we have denoted $\psi_{L}$ or $ \chi_{L}$, respectively.
\par
\begin{figure}[p]
\includegraphics[scale=0.5,angle=270]{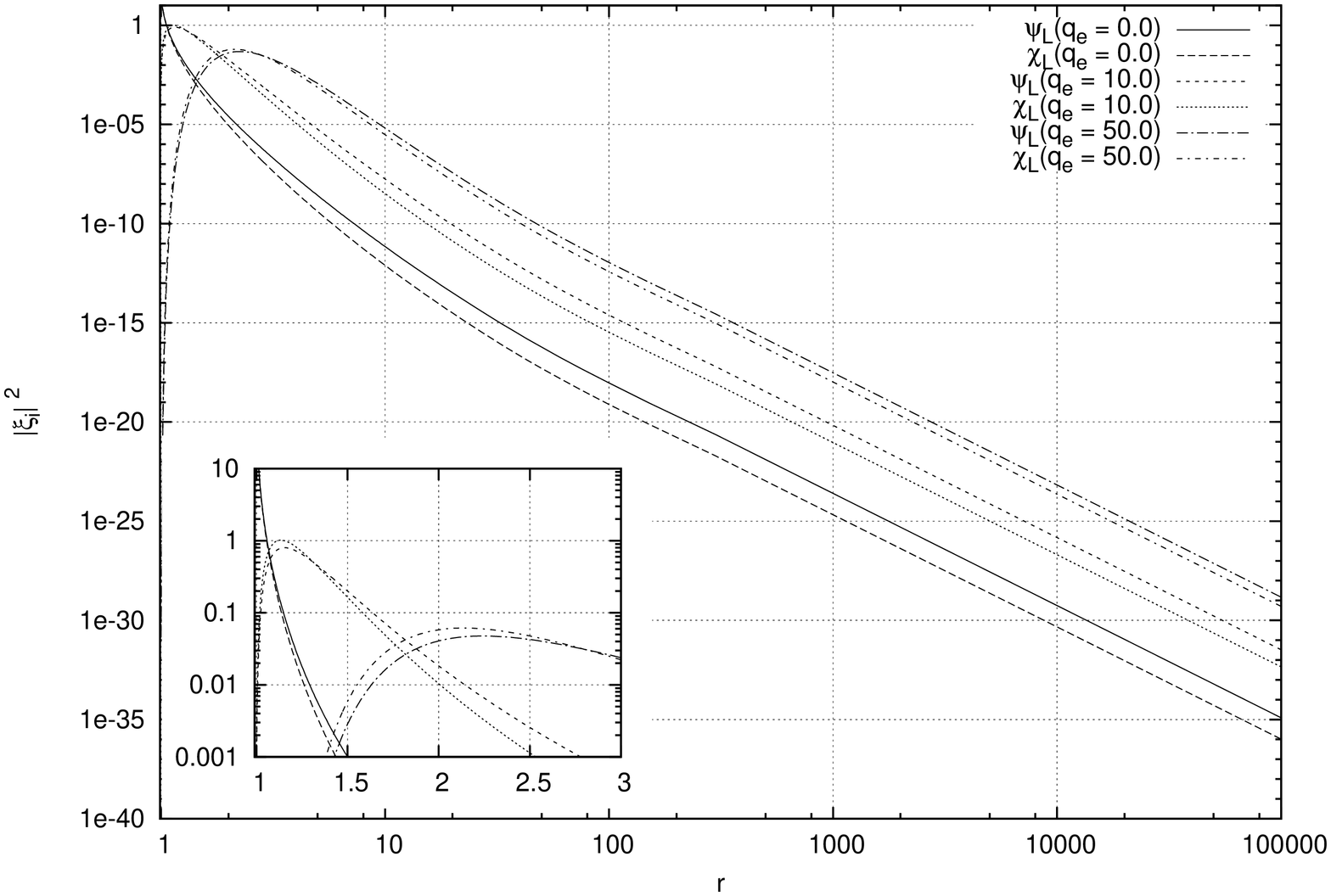}
\caption{Plot of fermion functions $|\xi_{i}|^2$, where $\xi_{i} = \{ \psi_{L},~\chi_{L} \}$ for the different 
values of the electrical charge in the background of nonextremal charged black string.
The other parameters are equal to $m_{f} = 2.7$, $N = 1$, $m = 1/2$, $q_{r} = 5$, $\omega = k = 0$.}
\label{fig1}
\end{figure}

In Fig.\ref{fig1} we plot $|\psi_{L}|^2$ and $|\chi_{L}|^2$ as a function of $r$-coordinate, for various values of the 
electric charge $q_e$. We set $q_e = 0.0,~10.0,~50.0$ and ${m}_{fer} = 2.7$, the winding number equals to $1$, the Higgs charge
$ q_r = 5$ and $m = 1/2,~\omega = 0,~k =0$. We shall first consider the case of non-extremal charged black string.\\
The solution with $q_e = 0$ is responsible for uncharged fermion field being the eigenstates of $\gamma^{0} \gamma^{3}$
matrices. The fermion functions $|\psi_{L}|^2$ and $|\chi_{L}|^2$ for the uncharged case are divergent near the black string
event horizon. On the contrary, the fermion function describing the charged fermions are regular in the nearby of the
aforementioned event horizon. The smaller $q_e$ we consider the closer the black string event horizon they begin to condensate.
\par 
Finally we note that at the beginning, when $q_e$ is small, fermions start to condensate just outside
the black string event horizon but inside the cosmic string core. These fermions are trapped as massless modes inside 
the Abelian Higgs vortex and they can lead to superconductivity. On the contrary, for larger $q_e$, the electrostatic
interactions among fermions and charge black string may eventually cause the expulsion of the fermions from the
considered cosmic string and destroy superconductivity.
One should mention that the electric charge has also a great influence on the width of the region where
fermion function $|\xi_i|^2$ values are different from zero
(let us say $|\xi_i|^2 > 10^{-10}$). Namely, the greater $q_{e}$ is the larger is the width in question.\\
For each electric charge,
one can find a specific value $r_e$ of the $r$-coordinate (for $q_e = 10, ~r_e =1.3 ,~q_e = 50, ~ r_e = 2.8$), where one has that
for $r > r_e$ the function $|\psi_{L}|^2$ has the greater values than $|\chi_{L}|^2$. On the other hand, when
$r < r_e$, the behaviour of the functions in question reverses.
\par
\begin{figure}[p]
\includegraphics[scale=0.5,angle=270]{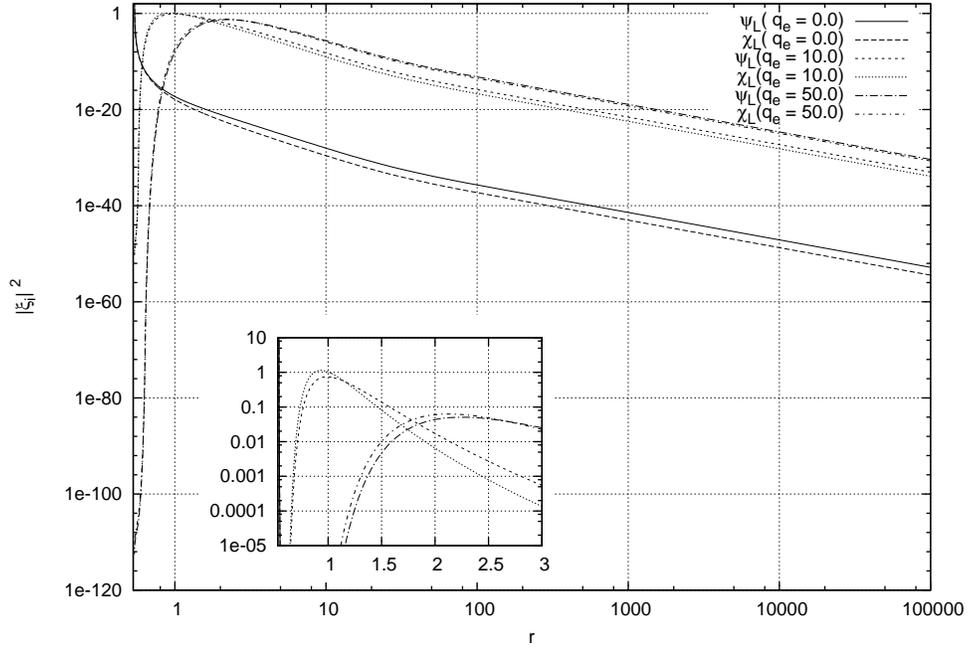}
\caption{Plot of fermion functions $|\xi_{i}|^2$, where $\xi_{i} = \{ \psi_{L},~\chi_{L} \}$ for the different 
values of the electrical charge in the background of extremal charged black string.
The other parameters as in Fig.1.}
\label{fig2}
\end{figure}

In Fig.\ref{fig2} we have elaborated the dependence of the fermion functions $|\psi_{L}|^2$ and $|\chi_{L}|^2$
on the electric charge for the extremal charged black string. We took into account the same values of the electric charge 
and other parameters as in 
Fig.\ref{fig1}. It turns out, that the uncharged fermion functions for which $q_e =0 $ are divergent 
near the extremal black string event horizon. On the other hand, the charged fermion functions are regular in the vicinity of it.
We also have the same dependence of the electric charge, i.e., the greater value of electric charge we have the farther 
from the event horizon of the extremal black string fermion fields begin to condensate. 
Comparing this effect in the spacetime of both types of black strings one remarks that
the extreme black string far more {\it expels} fermion fields that the nonextremal one. There is also 
the specific value $r_e$ ($q_e = 10,~r_e = 1.15,~q_e = 50,~r_e = 2.7$),
for which $r < r_e$ we acquire that $|\psi_{L}|^2 < |\psi_{L}|^2$ and $r > r_e$ function $|\psi_{L}|^2$ has 
greater values than $|\chi_{L}|^2$.
\par
\begin{figure}[p]
\includegraphics[scale=0.5,angle=270]{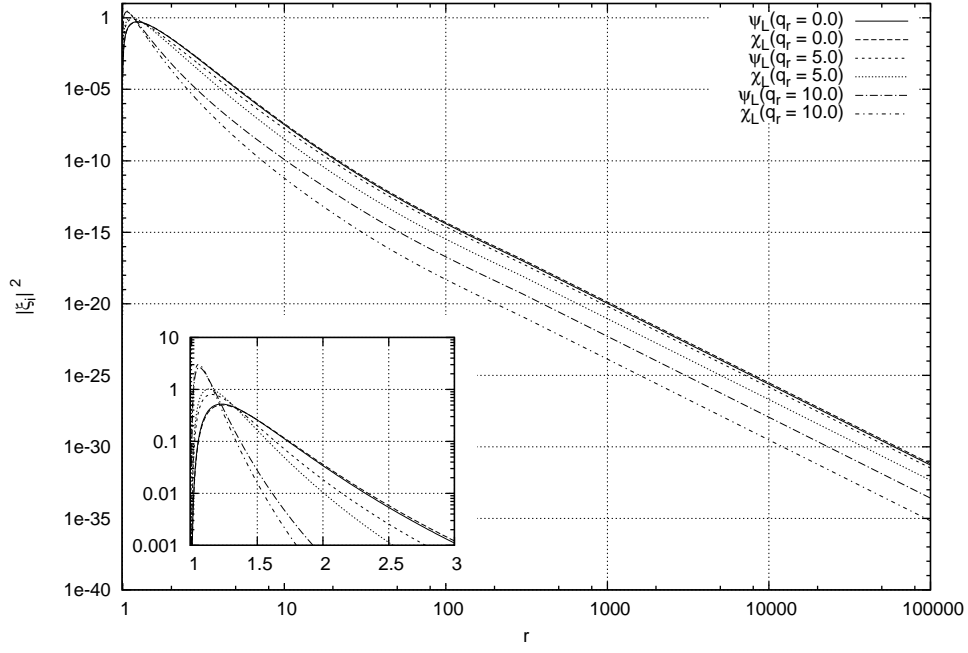}
\caption{Plot of fermion functions $|\xi_{i}|^2$, where $\xi_{i} = \{ \psi_{L},~\chi_{L} \}$ for the different 
values of the Higgs charge in the background of nonextremal charged black hole.
The other parameters we set $m_{f} = 2.7$, $N = 1$, $m = 1/2$, $q_{e} = 10$, $\omega = k = 0$.}
\label{fig3}
\end{figure}
\begin{figure}[p]
\includegraphics[scale=0.5,angle=270]{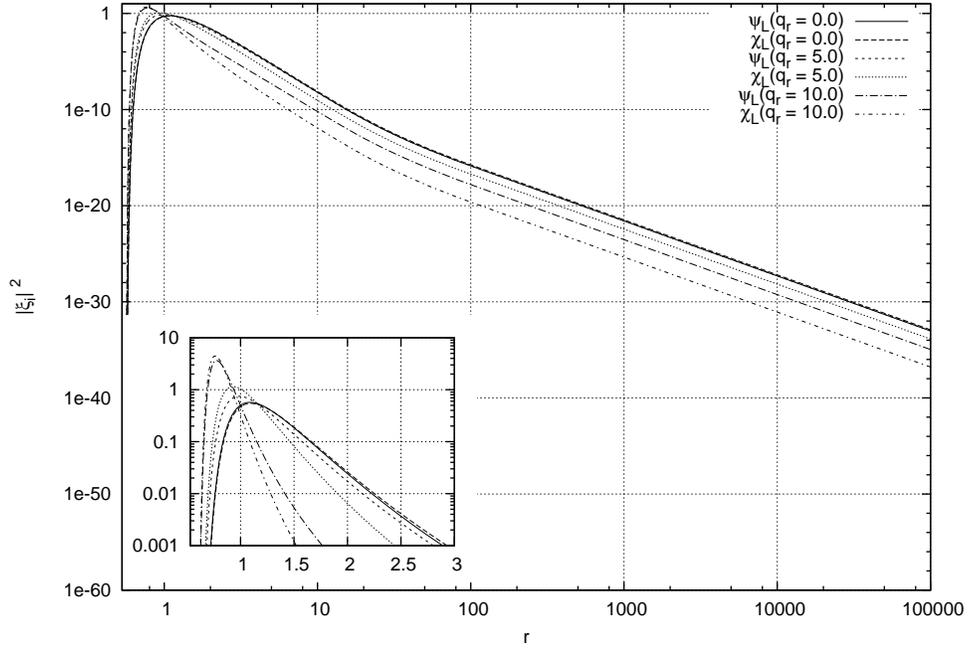}
\caption{Plot of fermion functions $|\xi_{i}|^2$, where $\xi_{i} = \{ \psi_{L},~\chi_{L} \}$ for the different 
values of the Higgs charge. The charged extremal black string case. The other parameters 
as in Fig.3.}
\label{fig4}
\end{figure}

In Fig.\ref{fig3} and Fig.\ref{fig4} we depicted the dependence of the fermion functions on the various values of the Higgs charge.
Namely, we take into account $q_r = 0.0,~5.0,~10.0$.
The electric charge was put to the constant and equaled to $10.0$. 
The other parameters are the same as in  Fig.\ref{fig1}. Fig.\ref{fig3} is valid for the non-extremal charged black string, while
Fig.\ref{fig4} is connected with the extremal case. 
The close inspection of the above figures reveals that $q_r$ has no influence on the regularity of 
the fermion solutions near the black string event horizon. For instance, for $q_{e} = 10$ and $q_{r} = 0$
the obtained solution is regular in the vicinity of the event horizon. However, the greater value of the 
Higgs charge one considers the closer to the black string event horizon fermions condensate.
For given value of the electric charge one has that the greater value of the Higgs charge we take into account the smaller
width of the region where $|\xi_i|^2$ is considerably different from zero and the larger maximal value of $|\xi_i|^2$
we obtain.
\par
\begin{figure}[p]
\includegraphics[scale=0.5,angle=270]{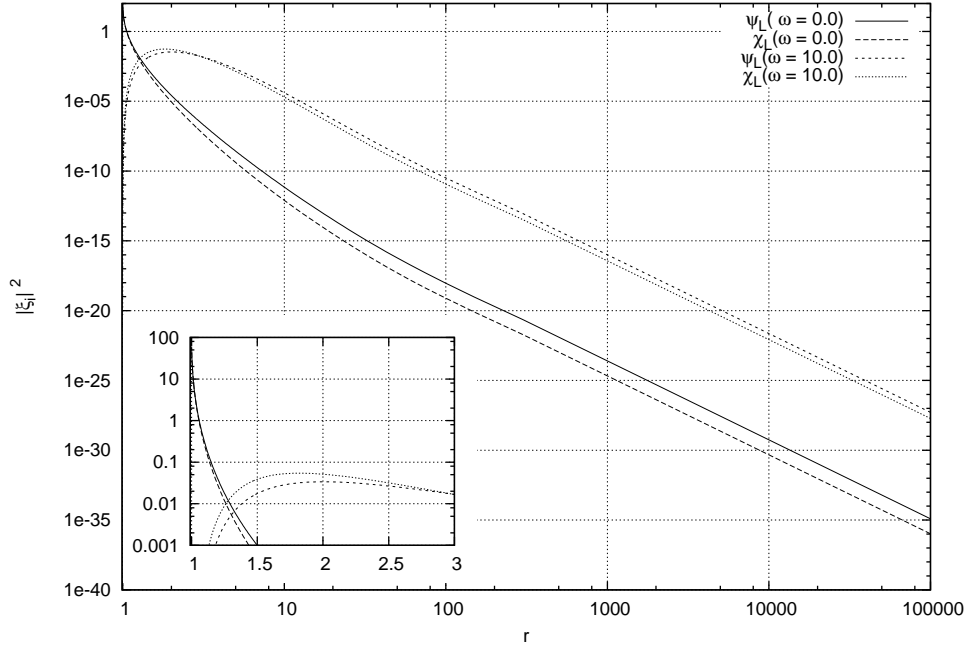}
\caption{Dependence of fermion functions $|\xi_{i}|^2$, where $\xi_{i} = \{ \psi_{L},~\chi_{L} \}$ on the different 
values of $\omega$. We set the winding number equal to $1$. The other parameters are 
are chosen to be
$m_{f} = 2.7$, $N = 1$, $m = 1/2$, $q_{r} = 5$, $q_{e} = 0$, $k = 0$. The nonextremal charged black string 
and electrically charged spinors case.}
\label{fig5}
\end{figure}
\begin{figure}[p]
\includegraphics[scale=0.5,angle=270]{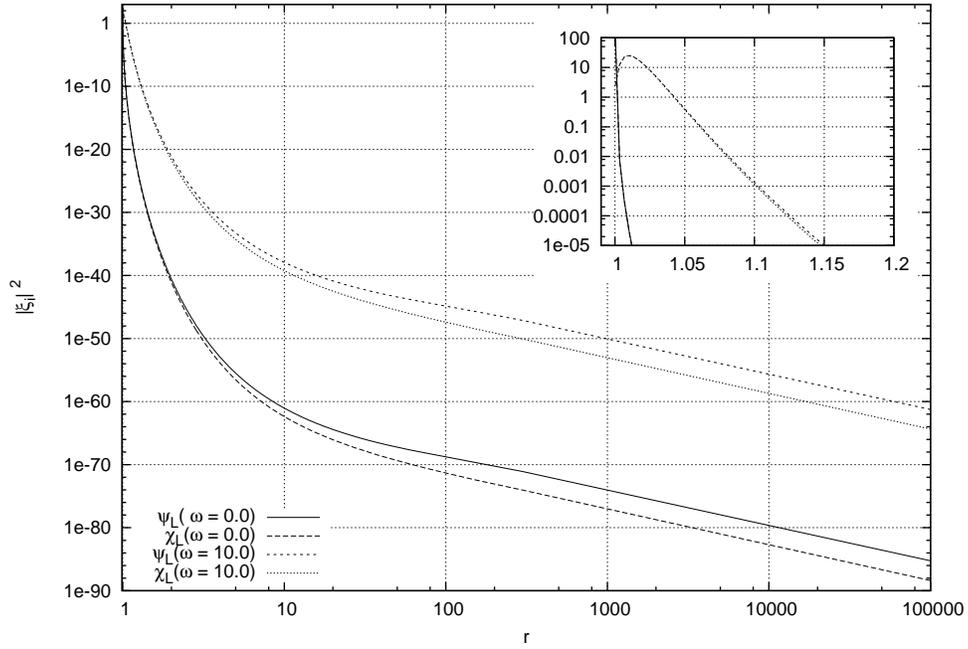}
\caption{Dependence of fermion functions $|\xi_{i}|^2$, where $\xi_{i} = \{ \psi_{L},~\chi_{L} \}$ on the different 
values of $\omega$. We put $N =10$. The other parameters are the same as in Fig.5. 
The nonextremal charged black string and electrically charged spinors case.}
\label{fig6}
\end{figure}
\begin{figure}[p]
\includegraphics[scale=0.5,angle=270]{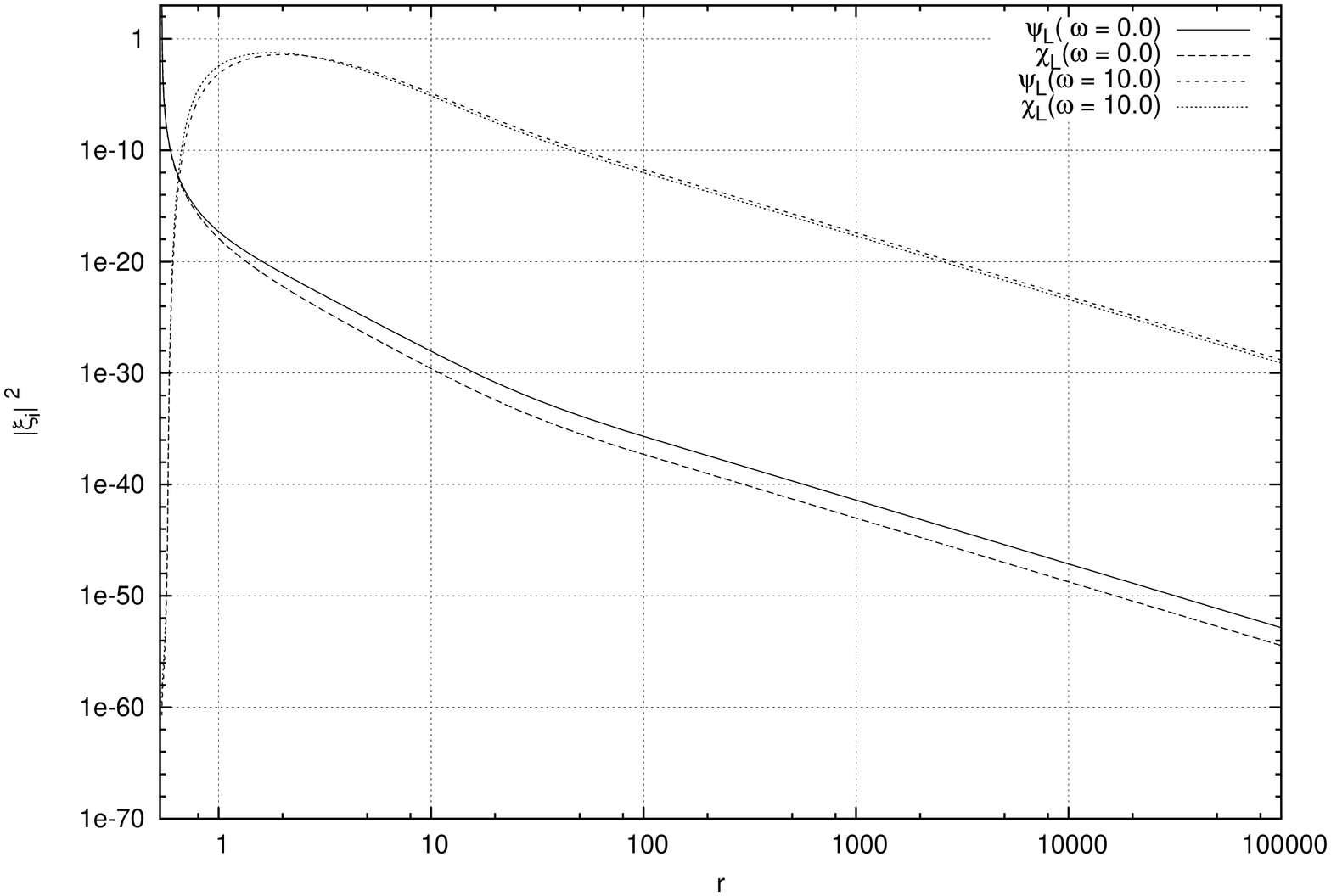}
\caption{Dependence of fermion functions $|\xi_{i}|^2$, where $\xi_{i} = \{ \psi_{L},~\chi_{L} \}$ on the different 
values of $\omega$. We choose the winding number $N=1$. The other parameters are the same as in Fig.5. 
The extremal charged black string and electrically charged spinors case.}

\label{fig7}
\end{figure}
\begin{figure}[p]
\includegraphics[scale=0.5,angle=270]{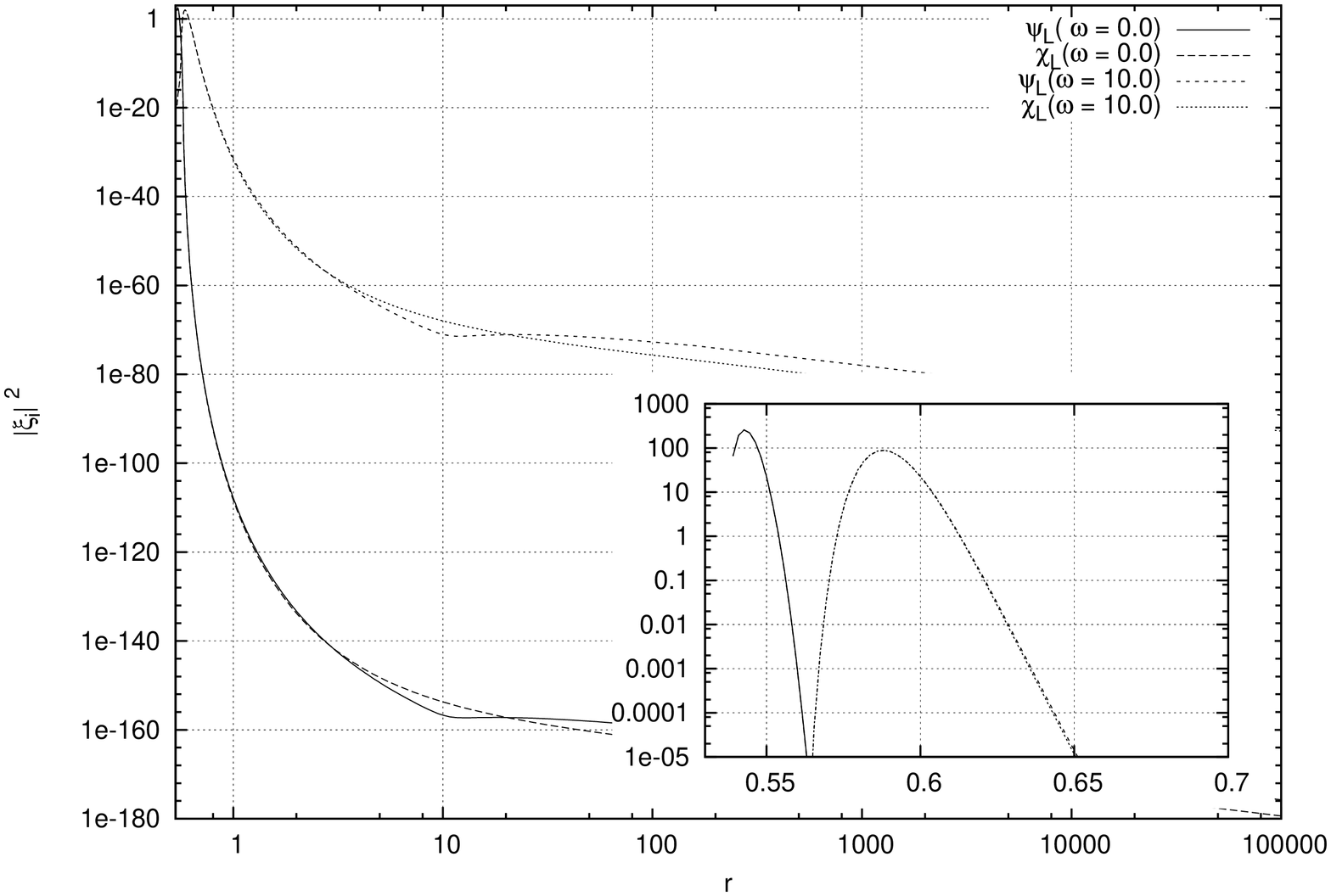}
\caption{Dependence of fermion functions $|\xi_{i}|^2$ where $\xi_{i} = \{ \psi_{L},~\chi_{L} \}$ on the 
different values of $\omega$ and the winding number $N = 10$.
The other parameters we set as in Fig.5. The extremal charged black string and electrically charged spinors case.}
\label{fig8}
\end{figure}

Now, we proceed to analyze the influence of the non-zero energy ($\omega \neq 0$) on the charged fermion functions.
In Fig.\ref{fig5} we study the nonextremal black string. The parameters we choose as
${m}_{fer} = 2.7$, the winding number $N = 1$, $m = 1/2$, the Higgs charge $q_r = 5$ and $q_e = 0,~k = 0$. 
We set $\omega = 0.0,~10.0$.
As we can see, even in the uncharged case, for the large enough $\omega$ we get solution regular 
in the nearby of the event horizon.
For $r > r_e =2.9$ one has that $|\psi_{L}|^2 > |\chi_{L}|^2$, but for 
$r < r_e$ the dependence reverses. In Fig.\ref{fig6} the parameters are the same as in Fig.\ref{fig5} 
but we consider the larger value of the winding number $N = 10$. Now, the larger value of the winding number caused 
that the localization of the fermion began closer to the black string event horizon. 
For $r > r_e = 1.07$ one has that $|\psi_{L}|^2 > |\chi_{L}|^2$, but when $r$ exceeds $r_e$ one arrives at the conclusion that
$|\chi_{L}|^2 > |\psi_{L}|^2$.\\
In Fig.\ref{fig7} and Fig.\ref{fig8} we take into account the same case of the non-zero modes for the extremal black string.
Namely, in Fig.\ref{fig7} the parameters are the same as in Fig.\ref{fig5} and we arrive at the regular solution with
$r_e = 1.2$. For the case when $N= 10$ one obtains that the curves depicting the behaviour of fermion functions intersect more than
one time and the closer value of $r_e$ to the black string event horizon is equal to $1.55$. When we consider 
the larger value of winding number we achieve the closer to the event horizon localization of fermion functions in question.
The other interesting feature is that for large N, even for $q_{e} = \omega = 0$, we get regular solution in the vicinity
of the event horizon.
\par
\begin{figure}[p]
\includegraphics[scale=0.5,angle=270]{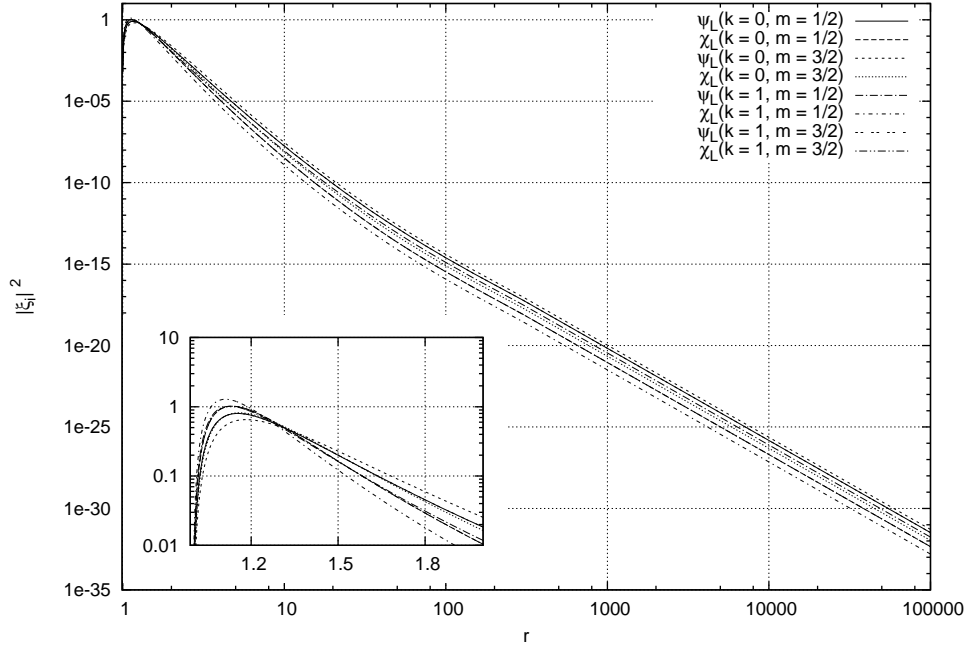}
\caption{Fermion functions $|\xi_{i}|^2$, where $\xi_{i} = \{ \psi_{L},~\chi_{L} \}$ for different values of $k$ and $m$
in the background of nonextremal charged black string. 
The other parameters are given by $m_{f} = 2.7$, $N = 1$, $q_{r} = 5$, $q_{e} = 10$, $\omega = 0$.}
\label{fig9}
\end{figure}
\begin{figure}[p]
\includegraphics[scale=0.5,angle=270]{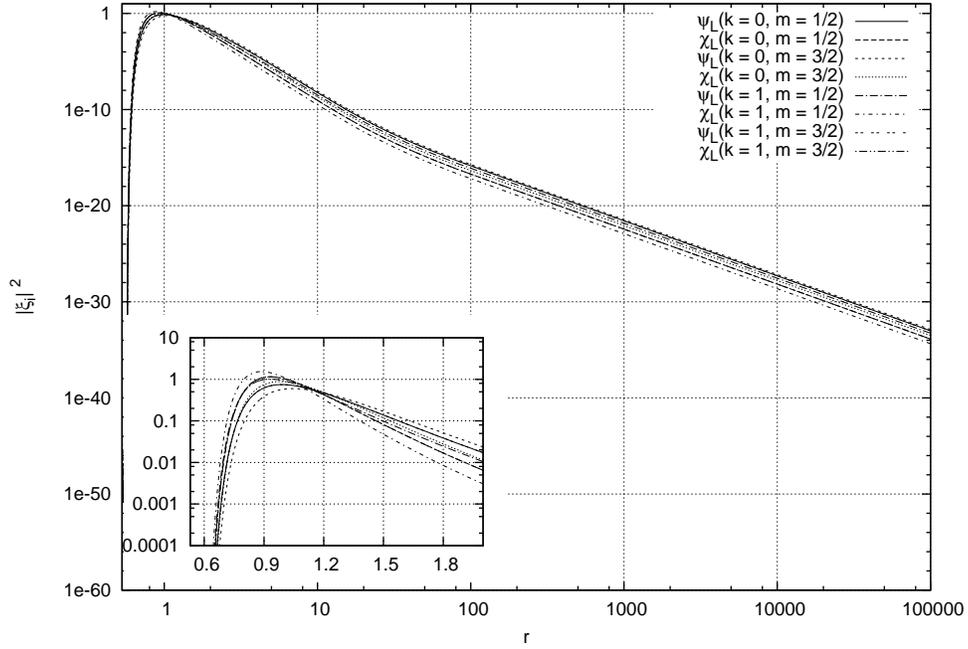}
\caption{Fermion functions $|\xi_{i}|^2$, where $\xi_{i} = \{ \psi_{L},~\chi_{L} \}$ for different values of $k$ and $m$
in the background of extremal charged black string. 
The rest of the parameters are set as in Fig.9.}
\label{fig10}
\end{figure}
\begin{figure}[p]
\includegraphics[scale=0.5,angle=270]{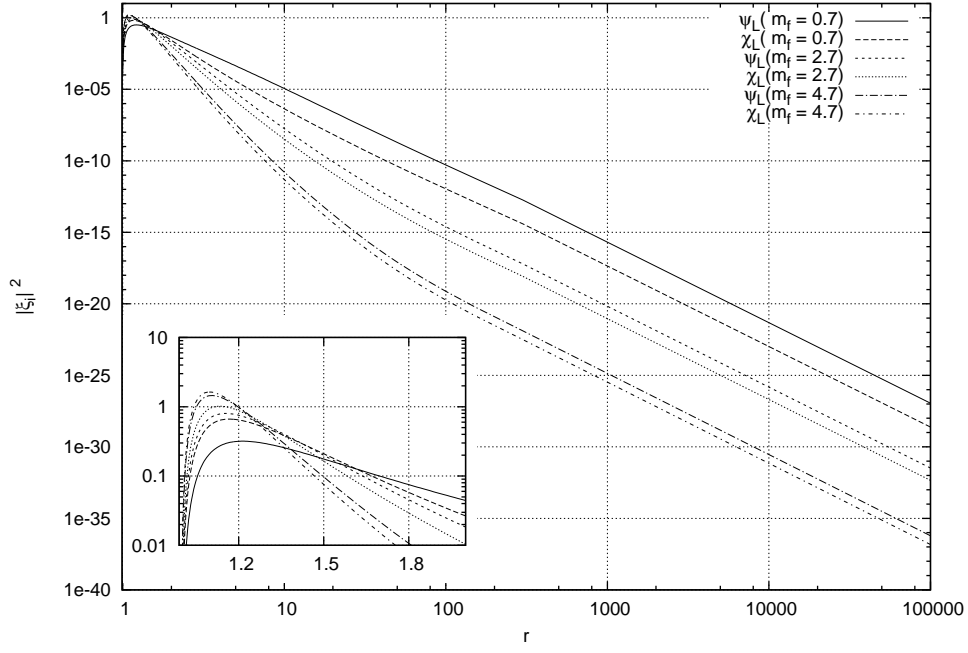}
\caption{Dependence of fermion functions $|\xi_{i}|^2$, where $\xi_{i} = \{ \psi_{L},~\chi_{L} \}$ on fermion mass for 
the nonextremal charged black string. 
The values of the parameters are: $N = 1$, $q_{r} = 5$, $q_{e} = 10$, $\omega =k = 0$.}
\label{fig11}
\end{figure}
\begin{figure}[p]
\includegraphics[scale=0.5,angle=270]{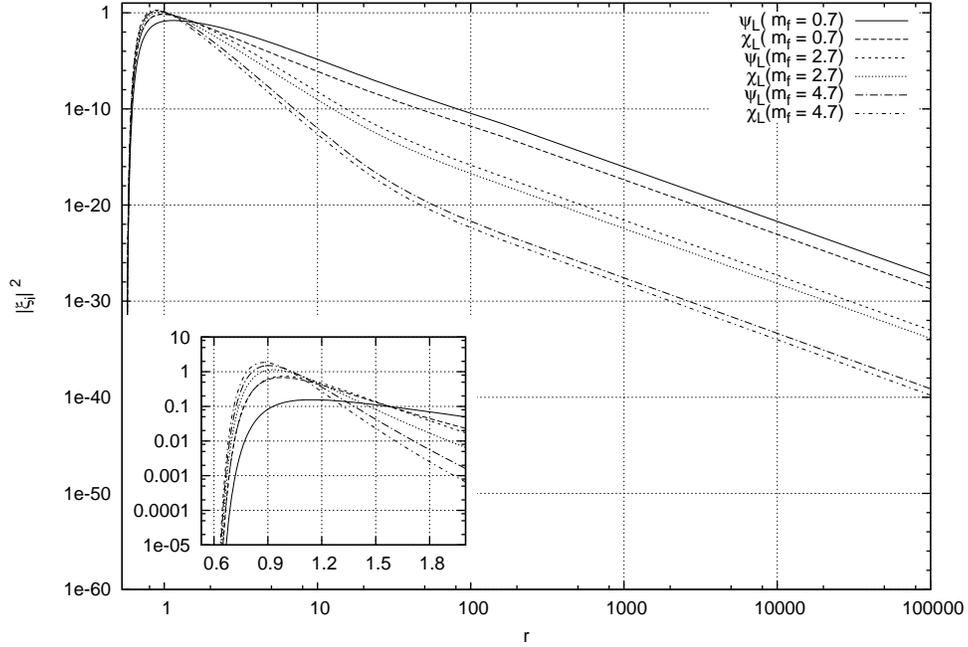}
\caption{Dependence of fermion functions $|\xi_{i}|^2$, where $\xi_{i} = \{ \psi_{L},~\chi_{L} \}$ on fermion mass for 
the extremal charged black string. 
The rest of the parameters are as in Fig.11.}
\label{fig12}
\end{figure}

In Fig.\ref{fig9} and Fig.\ref{fig10} we presented the behaviour of fermion functions for the different values of 
$k$ and $m$. The remaining parameters are the same as in the previous plots. 
One can conclude that near horizon of the extremal charged black string the fermion condensation
takes place farther comparing to the nonextremal black string.
\\
Fig.\ref{fig11} and Fig.\ref{fig12} are connected with the influence of the fermion mass ${m}_{fer}$ on the 
fermion functions in question.
We set ${m}_{fer} = 0.7,~~2.7,~4.7$
and the other parameters as in the previous cases. For each fermion mass, one attains that 
there is such a value $r_e$ for which one has that when $r > r_e$, then $|\psi_{L}|^2 > |\chi_{L}|^2$ and for $r < r_e$ we get 
$|\psi_{L}|^2 < |\chi_{L}|^2$. Moreover, the larger value of ${m}_{fer}$ is the smaller value of the fermion function one receives.
Near the charged black string event horizon the situation in question changes.
It turns out, that the bigger mass we have the larger
value is achieved by fermion function. The tendency that the extremal charged black string expels fermions 
more violently is maintained.

\section{Conclusions}
In our paper we have considered the problem of an Abelian Higgs vortex in the spacetime of a charged black 
string in the presence of Dirac fermion fields. Dirac fermions were coupled to the Abelian gauge fields $A_\mu$
and to the Abelian Higgs field $B_\mu$ as in the Witten's model of the superconducting cosmic string \cite{wit85}.
Moreover, we assume the complete separation of the degrees of freedom in the system in question.
One has studied the extremal  and nonextremal case of the black string pierced by an Abelian Higgs vortex. 
As far as the fermion function is concerned, we take into account the case of the uncharged, fermions being the eigenstates of 
$\ga^0~\ga^3$ gamma matrices, as well as the charged fermions. It was revealed that in the case of the uncharged fermions
we obtained the divergent solutions near the charged black string event horizon both in extremal and nonextremal cases.
On the contrary, the charged fermion functions are regular in the vicinity of the black string. The 
dependence of the fermion functions on the electric charge $q_e$ was elaborated. Namely, the smaller
 $q_e$ was, the closer to the event horizon fermions began to condensate. The same tendency was found in 
the case of the extremal charged black string. However, one remarks that the charged extremal black string expels
fermion fields far more violently than the nonextremal one.
\par
It worth mentioning that the
Higgs charge also plays the dominant role on the behaviour of fermion functions in the nearby of the black string event horizon. 
Namely, when we put $q_e$ equal to a constant value, it turned out that the greater value of the Higgs charge we considered
the closer to the event horizon fermion fields began to condensate. This was the case for both types of the black strings.
Nevertheless, for the nonextremal charged black string the condensation took place far more closer to the event horizon than 
in the case of the extremal black string.
\par
It is a remarkable fact that electric charge and Higgs charge are two parameters which have a great 
influence on the fermions in question. Especially, the fermion condensation depends on them. The increase of 
the electric charge provides the expulsion of fermions from the charged black string event horizon and eventually
even from the cosmic string core.
In turn, it can destroy superconductivity of the cosmic string in question, because of the lack of charge carriers inside the core.
Consequently, for large enough electric charge, instead of a superconducting cosmic string, one has an {\it onion-like}
structure. This structure consists of black string surrounded by cosmic string which in turn is encompassed by 
a shell of the fermionic condensate. Moreover, one has that for the larger value of the charge is taken into account
the larger width of the aforementioned shell one achieves. By the width of the shell in question we understand the region where 
$|\xi_i|^2$ are different from zero, e.g., $|\xi_i|^2 > 10^{-10}$.\\
Returning to the consideration of the Higgs charge, one can remark that the situation is totally different.
The increase of the Higgs charge implies the closer to the charged black string event horizon condensation of fermion fields
and the decrease of the width of fermion condensate shell.
\par
The winding number has also the influence on the behaviour of the considered fermion functions
$|\psi_{L}|^2$ and $  |\chi_{L}|^2$. For the established values of electric, Higgs charges, fermion mass, and for nonzero energy modes,
one obtains that the greater $N$ is the closer to the event horizon fermions begins to concentrate.
Fermion functions depend also on 
fermion mass $m_{fer}$. There is a point $r_e$ for which one has that, if $r > r_e$ then the smaller value of $m_{fer}$ one studies the 
the larger value of fermion function we attain. However, with the passage of $r$-coordinate 
in the direction to the event horizon, i.e.,
$r < r_e$, the situation alters.
\par
By virtue of the revealed features of the fermion functions in the background of a charged black string 
pierced by an Abelian Higgs vortex, one can draw a conclusion that in principle there is such a value of 
the electric charge which can destroy fermionic superconductivity. The winding number 
and Higgs charge also exert a great influence on 
the superconductivity carried by an Abelian Higgs vortex penetrating the black string in question. 
This is the case for both extremal and nonextremal charged black string vortex systems.


\begin{acknowledgments}
{\L}N was supported by Human Capital Programme of European Social Fund sponsored
by European Union. \\
MR was partially supported by the grant of the National Science Center
$2011/01/B/ST2/00408$.
\end{acknowledgments}


\end{document}